# A Reference Architecture for Autonomous Networks:
# An Agent-Based Approach

Joseph Sifakis[†], Dongming Li[*], Hairong Huang[*], Yong Zhang[*],
Wenshuan Dang[*], River Huang[*], Yijun Yu[*]

† *Verimag, Université Grenoble Alpes, Grenoble, France*
* *Huawei Technologies Co., Ltd., Bantian, Shenzhen, China*

*Abstract*

The vision of autonomous systems is becoming increasingly important in many application areas, where the aim is to replace humans with agents. These include autonomous vehicles and other agents' applications in business processes and problem-solving. For networks, the increasing scale and operation and management (O&M) complexity drive the need for autonomous networks (AN). The technical objective of AN is to ensure trustworthy O&M without human intervention for higher efficiency and lower operating costs. However, realizing AN seems more difficult than autonomous vehicles. It encounters challenges of networks' structural and functional complexity, which operate as distributed dynamic systems governed by various technical and economic constraints.

A key problem lies in formulating a rigorous development methodology that facilitates a seamless transition from traditional networks to AN. Central to this methodology is the definition of a reference architecture for network agents, which specifies the required functionalities for their realization, regardless of implementation choices. This article proposes a reference architecture characterizing main functional features, illustrating its application with network use cases. It shows how artificial intelligence components can be used to implement the required functionality and its coordination. The latter is achieved through the management and generation of shared domain-specific knowledge stored in long-term memory, ensuring the overall consistency of decisions and their execution. The article concludes with a discussion of architecture specialization for building network layer agents. It also identifies the main technical challenges ahead, such as satisfying essential requirements at development or runtime, as well as the issue of coordinating agents to achieve collective intelligence in meeting overall network goals.

# 1  Autonomous Networks – Vision and Challenges

## 1.1  Autonomous System Development Trend

### 1.1.1  Challenges and prospects for the autonomous systems

Autonomous systems arise out of the need to automate existing organizations by progressive and incremental replacement of human operators by autonomous agents. These are complex distributed systems made up of agents each pursuing specific goals, that must work together to achieve the system's overall goals. They are fundamentally different from game-playing robots or intelligent personal assistants. They must fulfil critical properties and demonstrate "broad intelligence" by handling knowledge to adapt to unpredictable and complex environments. This implies the ability to manage sets of goals that evolve dynamically and are potentially conflicting.

The development of trustworthy autonomous systems, as anticipated by the Industrial Internet of Things (IIoT), is considered a bold step toward closing the gap between human and artificial intelligence. However, reaching this vision raises fundamental questions that remain poorly understood, including the need for agreement on a concept of autonomy that is consistent with the current vision of autonomous systems. While extensive research has been conducted on autonomous agent systems and multi-agent systems within the field of AI over several decades, much of this work has focused on agents operating in predictable digital environments. Autonomic computing, for instance, had its heyday in research at the beginning of this century, powerfully promoted by IBM [1, 2]. However, it was focused on agents that are very different from agents designed to replace humans in performing tasks in a socio-technical environment.

The development of autonomous agents must go beyond logical aspects. It must also address risk analysis, mitigation, and evaluation, focusing on the hazardous situations that can result from the interaction of the agent with its environment. Existing techniques fall short because they assume an exhaustive analysis of the causes of the risk and the deployment of mitigation mechanisms at the design stage. These techniques are impracticable due to the inherent complexity of autonomous system environments. One trend to overcome these difficulties is to move away from the idea of correction at design time, and explore emerging ideas that rely on run-time assurance techniques [3] replacing static mitigation mechanisms.

The global validation of autonomous systems, regarded as ensembles of autonomous agents interacting to achieve collective goals, challenges rigorous validation techniques due to complexity issues and the lack of adequate modelling frameworks. Moreover, complexity is aggravated by parameters like the number of lines of code or components and the temporal and spatial dynamism of agent interactions with the cyber-physical and human environments.

Simulation and testing remain the only feasible approach to assess the overall trustworthiness of an autonomous system. However, existing theories and techniques for testing hardware and software systems are model-based. They pursue well-defined objectives in the form of coverage criteria or achievement of test purposes that can be formulated and understood rigorously. Their application to autonomous systems requires the development of far more powerful modelling and testing techniques, as well as the formalization of adequate success criteria.

Many believe that we need to break with current system development techniques, which are a barrier to the acceptance of new technologies. Current trends in autonomous systems are overturning traditional engineering practices. Automatic updates of critical software of autonomous vehicles violate the rules according to which a given product, once developed and certified according to standards, must not undergo modifications. Furthermore, in some countries, such as the USA, self-certification is permitted for autonomous vehicles, leaving manufacturers with the responsibility of guaranteeing their safety.

Systems engineering is reaching a turning point, moving from small, automated, centralized, non-scalable systems with predictable environments, to large, autonomous, distributed, reconfigurable systems with unpredictable, dynamically changing environments. The two main problems, which are of comparable importance, are 1) the design of autonomous agents that can pursue predefined goals in unpredictable environments, and 2) the global validation of autonomous systems composed of an arbitrary and dynamically changing number of interacting agents.

### 1.1.2 Characterization and classification of autonomous systems

The aim of autonomous systems to replace human agents implies that they combine features that make them the most difficult systems to realize. Figure 1 classifies the different types of systems according to two criteria. The first criterion is the way they interact with their environment.

➢ **Transformers** have the simplest input/output behavior: for a given input they provide a corresponding output. These can be simple functions or conversational systems such as ChatGPT.

➢ **Predictors and Analyzers** are streamers that receive sequences of data and produce knowledge about a monitored environment. So, their current output depends on the history of the events received. They find numerous scientific and industrial applications such as weather prediction, analysis of critical situations and root cause analysis.

➢ **Controllers** are systems that continuously interact with their environment in a closed loop so that the observed global behavior satisfies desirable properties. Their complexity can vary depending on the type of their environment and also the type of the properties to meet. An important distinction lies in the possibility to figure out all the possible interactions between the controller and its environment and build a "static controller". Controllers of traditional automated systems such as lifts, thermostats or flight controllers, are static. On the contrary, for a chess playing system it is practically impossible to build a static controller. We need a dynamic controller that adopts adaptive behavior according to the opponent's movements. The behavior of dynamic controllers is a two-player game between them and their environment. It boils down to computing policies pursuing a set of goals on a predictive model that is a transition system showing possible actions of the controller to counter reactions of its environment.

Another criterion for system classification is the distinction between technical and non-technical systems.

➢ **Technical systems** are systems whose I/O relation can be unambiguously characterized as a relationship between mathematical domains. Traditional ICT systems is technical systems. Their trustworthiness can be formulated using testable specifications which are predicates linking sequences of input values to sequences of output values. For these systems, it is possible to apply verification techniques that compare a model of their behavior against their specifications. Their correctness can be objectively formulated and their reliability assessed. The risk induced by their use is regulated by standards that require reliability commensurate with their criticality. There are, however, technical systems whose behavior is difficult to specify rigorously and can only be tested.

➤ **Non-technical systems** are systems whose input or output domains involve data requiring interpretation, such as sensory or linguistic data. These are systems that mimic human functions that do not lend themselves to formalization, e.g. ChatGPT, or image analysis systems. In addition to safety, security and performance, their trustworthiness is characterized by human-centered properties that describe their degree of alignment with human values. Their correctness cannot therefore be validated objectively. It is worth noting that AI systems open up a new era in systems engineering by enabling non-technical systems to be built. However, AI systems can also be used to build technical systems whose intrinsic complexity defeats traditional model-based methods. For example, a chess playing system or a weather forecasting system are technical systems for which AI solutions are more practical.

Figure 1, shows the above classification of systems depending on the way they interact with their environment and the formal specification of their I/O relation. Autonomous systems, in their most general acceptation, are non-technical dynamic control systems.

| Type of System / I/O relation | Transformer | Predictor | Analyzer | Controller | |
|---|---|---|---|---|---|
| | | | | Static | Dynamic |
| | | TRADITIONAL SYSTEMS | | | |
| **Technical system** (Mathematically well defined I/O relation) | Functions, Combinatorial circuits | Predictive statistical models, e.g. weather forecast | Static analyzers, Model-checker Symbolic solver | Thermostat, Lift controller, Flight controller | Chess playing system, MPC system, Dynamic decision making. |
| **Non technical system** (Involves Linguistic or Sensory data) | Conversational system, Classifiers. | Regression models, Time series analysis, Decision trees. | Root cause analysis Clustering system. | Language-controlled manipulation, Object searching in a space. | Collaborative agent, Football robot, Self-driving autopilot, Drones |
| | | AI SYSTEMS | | | AUTONOMOUS SYSTEMS |

Figure 1 The status of autonomous systems in relation to the distinction between technical and non-technical systems

## 1.1.3 Adoption of autonomous systems by industry

As explained, autonomous systems are essential for achieving the IIoT vision considering their application in transport systems, communication networks, smart production systems, smart farms, and various services. Their realization requires a marriage between traditional systems engineering and AI that relies on solid foundations to guarantee their trustworthiness.

A well-known fact about AI systems is that they are black-boxes and it is practically impossible to explain their behavior using mathematical models, as we can do for traditional systems. Their development remains empirical and does not follow the traditional approach that builds a system's architecture by progressive and reasoned analysis of requirements. Traditional critical system development relies on models to guarantee their reliability, by application of verification and analysis techniques. Traditional test methodologies cannot be applied to non-technical AI systems, such as ChatGPT, as it is impossible to build an oracle that can judge based on objective criteria. But even for technical AI systems, we do not have test methods based on coverage criteria such as those we apply to hardware or software systems. The fact that they are not explainable and therefore the absence of behavioral models, limits their empirical validation to statistical testing. While it is conceivable to use AI systems in conversational mode with humans in the loop, it is unthinkable to integrate them into autonomous industrial systems, whose high criticality imposes trust requirements of a comparable level.

These inherent limitations of AI systems are not the only obstacle on the road to industrial autonomous systems. An autonomous system can be made up of a large number of agents that need to coordinate to achieve global systems goals. This raises challenging systems engineering problems that characterize the coordination of real-time distributed systems. In particular, autonomous systems are subject to the need to adapt to unpredictable environments and must be able to reconfigure themselves dynamically by creating or deleting agents and possibly changing their architecture.

Autonomous vehicles are a topical case, emblematically illustrating the difficulties in moving from automation to autonomy. Despite massive investment and the enthusiastic commitment of the sector's major players, who promised total mobile autonomy by 2020, the autonomous car industry is struggling to turn this ambition into reality. Today, the most advanced projects are limited to the deployment of robotaxis in urban areas [4, 5], with limited success. In addition to sporadic safety problems, they also lack the social intelligence of human drivers, e.g., not giving way to priority cars, or stopping unexpectedly [6, 7].

Currently, there are two different technical approaches for developing such systems, but both fall short of addressing the full autonomy challenge.

➢ **_Modular Construction_**. One approach adopts a modular construction in which an agent is the composition of building blocks each implementing a function, so that the resulting overall behavior meets the essential requirements for autonomy. The solutions, also known as hybrid solutions, can integrate data-based components, e.g. for perception, and model-based components, e.g. for decision-making. Modular construction allows understanding and to some extent explainability of the agent behavior by applying compositionality principles. It is adopted by some ADS manufacturers such as Baidu/Apollo and Waymo following the traditional systems engineering methods and applying as much as possible existing techniques to guarantee trustworthiness. This approach however, may suffer limitations due to the complexity of the overall construction and in some cases the lack of composability of the composed features.

➢ **_End-to-end AI_**. Recently, interest has shifted from traditional modular autopilots to end-to-end AI autopilots for two reasons. Firstly, development costs are lower for end-to-end solutions. Secondly, the adoption of a holistic development minimizes the risks inherent in the non-composability of functionality in modular architectures. However, end-to-end AI autopilots are less deterministic and suffer from the well-known problems of AI systems, such as non-explainability and anomalies [8, 9, 10]. Such end-to-end solutions adopted by some autonomous driving industry, take sensory data from the vehicle's various sensors and produce a trajectory in the form of a timed sequence of vehicle states, realized by a dedicated controller or converted directly into steering angle and acceleration/deceleration signals. They are developed by training neural networks with data either collected by vehicles or synthetic data generated by simulators.

The problems encountered by the automotive industry are found to varying degrees in other sectors currently less advanced on the road to autonomy. In other transport applications, such as avionics, autonomy is being applied mainly to drones. Autonomous applications in production and services are also progressing satisfactorily, with the increased use of robots. Of all the application areas, autonomous communication networks appear to be an extremely challenging case. Except for perception issues, the problems posed by autonomous networks (AN) are infinitely more difficult than those posed by autonomous cars. From the outset, AN pose the issue of collective intelligence. It is not a question of creating a single autonomous agent, but of putting intelligence at all levels of an extremely complex, dynamic, and reconfigurable hierarchical structure. This means that a large number of agents must coordinate to achieve global goals at the business and network levels while fulfilling their specific goals autonomously. A central question in this case is how to transfer the knowledge of thousands of operation and maintenance (O&M) engineers to AI and knowledge management systems.

For AN, end-to-end AI solutions can be applied to network components, but it is unthinkable to apply them to an entire network, where decisions of different kinds have to be made at different levels and in a distributed way. It will therefore be inevitable to use "hybrid" solutions integrating different types of AI and ICT components.

Because of their scale and complexity, AN pose problems in terms of development methodology. Traditional methods based on the Vee model [11] do not apply to networks that adopt an architecture-based approach from the outset. This approach is based on the idea that architecture, as a mechanism for coordinating integrated components, can guarantee certain global properties by construction. The term reference architecture is used to designate a generic solution to a coordination problem. Many de facto standards are based on reference architectures, which are generally open and can be tested and evaluated by an entire community. Their importance in all areas of complex systems engineering is demonstrated by their numerous applications.

The first step in applying the architecture-based design paradigm to AN is to define an agent reference architecture. Such an architecture must be sufficiently general to encompass all its specific uses in different contexts and layers of a network.

The next steps should aim to progressively integrate agents into the overall network architecture for sustainable and profitable operation of the autonomous network through interoperability, connectivity, and distributed intelligence. In particular, the aim is to achieve global situational awareness to offer optimization and value-creation services.

## 1.2   Vision and Direction of the Autonomous Networks

### 1.2.1   Why we need autonomous networks?

The need for autonomous networks (AN) arises from various industries, carriers, operation and maintenance (O&M) engineers, and even end users. It raises three challenges posed by increasingly complex networks and the demand for greater reliability and efficiency.

***Overcoming the rapid increase in network scale and complexity***. Large-scale deployments and the integration of IoT, cloud services, and 5G have made modern networks more intricate than ever. The accelerated growth of technologies such as AI, augmented and virtual reality, blockchain and decentralized systems, as well as the Internet of Things (IoT), will lead to increasingly large and complex networks in the future. Traditional manual O&M approaches struggle to effectively manage this level of complexity, as they require substantial human intervention and specialized expertise.

***Minimizing reliance on human intervention in network operations***. Most networks are still primarily operated and managed by O&M engineers for configuration orchestration, service provisioning, experience or performance optimization, etc. This poses several problems for the future development of networks. First, O&M engineers are required to have high expertise, skills, and qualifications, and deep familiarity with network services, topology, configurations, and operational proficiency. Furthermore, this requires large O&M teams, which adds considerably to labour costs. Additionally, errors such as misoperations or review oversights are inevitable, which will introduce vulnerabilities and potential faults into the network. From the user's perspective, certain tasks, such as network recovery or service activation, require additional involvement on the part of the user, leading to a less satisfactory experience.

***Efficiently and intelligently healing the network***. Outages disrupt the network for businesses and end users, significantly impacting the user experience. It is difficult to isolate faults before they cause damage due to the lack of effective automatic detection or prediction techniques. Once a failure occurs, addressing the issue (through identification, localization, analysis, and resolution) requires significant effort from O&M personnel. What's more, the process takes a long time, which means that users must wait a long time before the network is restored. Consequently, improving the capacity for healing through AN is crucial to maintaining the reliability of the network and minimizing service interruptions.

In addition to the factors mentioned above, other considerations, such as faster service provisioning and the application of data-driven methods are also important. The demand for AN is growing rapidly, which is what is driving our in-depth research on the subject.

### 1.2.2    The vision for autonomous networks

***Global Vision***. We present the overall technical vision of AN, which is developing the future network system based on agents, called the AN agent. It will replace humans in operation and maintenance tasks. We require that the AN agent be capable of providing services covering the following aspects:

➢ Sense its environment and adapt its behavior based on acquired knowledge to respond to environmental stimuli.

➢ Guarantee network trustworthiness, characterized by sets of properties that imply compliance with user and operational requirements.

***Self-\* and Zero-\* vision***. The AN vision has been proposed in [12, 13] , which can be examined from two perspectives: how the O&M engineers interact with the network and how end-users experience the network.

From the O&M perspective, three key self* requirements guide the development of the AN agents:

➢ *Self-Fulfilling (Self-Organizing and Self-Configuring)*: The AN agents are capable of autonomously provisioning, orchestrating, coordinating, and configuring network services or resources by accurately understanding and identifying the user's initial requirements.

➢ *Self-Healing*: The AN agents possess the ability to precisely detect, identify, and locate faults, enabling them to autonomously address issues and prevent service degradation or outages, thereby ensuring continuous high availability and reliability of network services.

➢ *Self-Optimizing*: The AN agents can autonomously adjust configurable parameters to optimize network performance and enhance user experience, thereby ensuring the continuous delivery of high-quality network services.

From the perspective of network users, three key zero* requirements define the experience created for end-users through the use of AN agents:

➢ *Zero-Wait*: The measured waiting or delay duration perceived by users during service provisioning, orchestration, coordination, and configuration is minimized, ideally approaching zero.

➢ *Zero-Touch*: The measured frequency and number of network service operations required by users or O&M engineers are substantially reduced, ideally with zero direct intervention.

➢ *Zero-Trouble*: The measured duration of network service degradation, outage occurrences, failure rates, and customer complaint rates are significantly minimized, ideally approaching zero.

Note that all these requirements must be broken down into sets of properties that must be satisfied by the various components of the autonomous network.

### 1.2.3   Where we are today

Most complex analysis and decision-making tasks are currently performed by humans, requiring significant time, effort, expertise, and skills from O&M engineers. However, as network complexity and the diversity of use cases continue to grow, the manual O&M approach is encountering increasingly significant bottlenecks, which highlights the need for autonomy.

For example, IP network changes represent a typical O&M use case, frequently occurring in data centre networks (DCN) and metropolitan area networks (MAN). These changes involve tasks such as device software version upgrades, patch updates, hardware replacements, route configuration, and network structure adjustments. The primary challenge lies in ensuring that the configuration command scripts are accurate, meet the requirements of the network changes, and do not disrupt network services. Currently, this process is performed manually by humans and involves the following steps:

1. Change Solution Formulation: Based on requirements, the O&M engineers devise a solution that includes configuration scripts and procedural steps. This step relies heavily on the personnel's network knowledge and experience, which introduces risks of errors.

2. Solution Review: The solution undergoes a review by O&M engineers to identify potential risks or errors. However, as this step also depends largely on empirical expertise, potential errors may occasionally be overlooked, resulting the network faults.

3. Implementation on the network: Once approved, the solution will be implemented on the live network. This step could be vulnerable to human operational errors, leading to network incidents.

From the use case of network changes and numerous other examples, it is evident that the current approach heavily relies on human involvement in various O&M tasks. This over-reliance leads to high costs and low efficiency. Additionally, the network is at risk of disruptions caused by inadvertent human errors.

### 1.2.4 Where we go – future direction

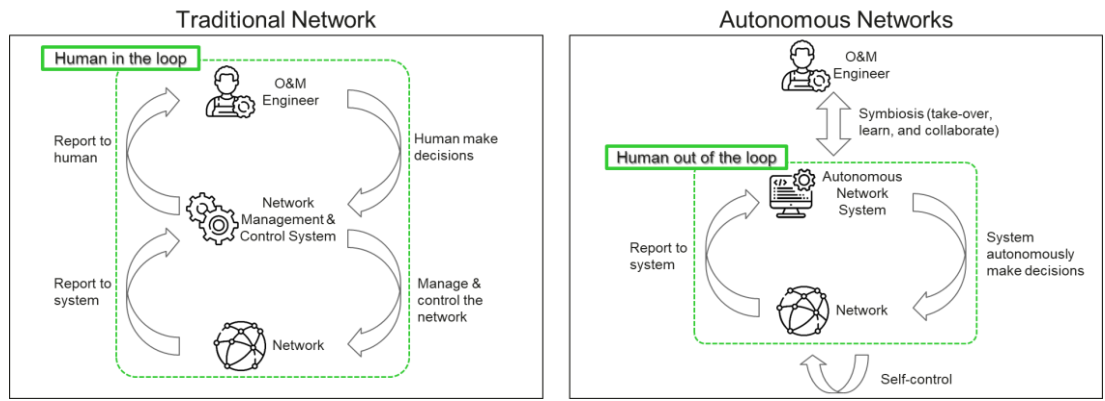

Figure 2 The transition from current to autonomous networks and the respective role of O&M engineers

The transition from the current networks to AN will take a long time. From the O&M perspective, the major difference is shifting from human-in-the-loop to human-out-of-the-loop, as seen in Figure 2. Unlike current networks that require manual intervention in configuration, optimization, and troubleshooting, AN operate with minimal or no human input. They should enable the network's O&M engineers to focus on strategy, goal, improving intentions and providing knowledge, rather than on day-to-day maintenance.

Compared to autonomous systems in other industries, such as autonomous vehicles and smart factories, future systems for AN probably are the most difficult to develop, operate, and maintain due to multiple factors:

***System Complexity***. AN are highly dynamic and constantly evolving online, with a reconfigurable structure and features capable of operating without interruption. They must also adapt to constantly changing environments and user demands, which requires supplementing design-time correctness with runtime correctness. Furthermore, these networks are highly distributed, demanding real-time operational efficiency.

***Layered Network Structure***. AN are made up of multiple interconnected systems across Business, Service, and Resource layers, necessitating efficient cross-layer communication and coordination. Additionally, the network infrastructure is highly heterogeneous, comprising a diverse array of components such as sensors, network devices, servers, and application software.

Given the scale and difficulty of the AN undertaking, a collective effort will be required to make the AN vision a reality. We believe that the following work needs to be done.

➤ Above all, we need to define a **functional reference architecture** for the AN in a rigorous manner, which includes basic architectural principles, behavior supported by functional characterizations, and internal and external interactions. The architecture must be a general pattern followed to guide the instantiation of agents and used for the comparison between different agent solutions.

➤ Following the examples of other sectors, such as avionics automotive, production automation, we should develop a rigorous **conceptual framework** for specifying and validating networks to allow achieving a common understanding of the problems and to build adequate technical solutions.

➤ In parallel, we need to **develop infrastructure and AI techniques**, particularly for AN, including but not limited to 1) collecting and cleaning telecommunication data for different types of AI systems and setting up the associated data infrastructure; 2) building AI components for prediction and analysis; 3) developing telecom foundation model and domain-specific model or algorithm; 4) transferring human expertise in O&M to a knowledge management system.

➢ Finally, a considerable effort should be made to **develop libraries** for components, architectures and protocols as well as **tools** for development, validation, and management.

In this paper, we propose a reference architecture for autonomous network agents characterizing key functionalities (Section 2). In addition, we analyze current challenges and propose lines of work for the development of the techniques essential to the realization of AN networks (Section 3).

# 2 Reference Architecture for Autonomous Networks Agents

This section presents a reference architecture for agents applied in autonomous networks (AN). It is structured as follows.

In Section 2.1, we present the objective and principles for designing this architecture.

In Section 2.2, we present the reference architecture for an AN agent. A brief view of the architecture with its interaction flows is provided in Section 2.2.1. Sections 2.2.2 and 2.2.3 define two complementary types of AN agent behavior. Reactive behavior concerns the interaction between the agent and the network environment it controls. Proactive behavior concerns the agent's intention to face risks and adapt its behavior in order to fulfill its purpose. We explain how these behaviors are achieved by composing modules that perform essential functions, and how they can be applied to real network use cases (Section 0). Sections 2.2.4 and 2.2.5 detail the importance and functionality of the Agent-Agent Interaction and Human-Agent Interaction modules, which deal with interactions with other agents and human operators respectively. In Section 2.2.6, we describe the World Knowledge (WK) module, which is a knowledge repository containing all the essential knowledge required to support the functionality of the other modules.

In Section 2.3, we outline the hierarchical architecture of AN, and its instantiation guidance as an agent instance in the hierarchy.

In Section 2.4, we define the criteria for classifying network operation and maintenance tasks according to their reactive or proactive role, and provide examples of their application in real network use cases.

In Section 2.5, we present the achievement incrementality of the architecture with three progressive phases.

## 2.1 Architecture design objectives and principles

The objectives of defining the AN agent reference architecture include:

➢ To provide a general **architectural pattern** to guide the development and implementation of an AN agent instance, so that the realized AN agent can replace humans in operation and maintenance (O&M).

➢ To explicitly and rigorously specify the **functional behavior** of an AN agent, as a composition of modules with well-defined interfaces, and interactions.

The definition of the architecture is guided by the following general principles:

➢ ***Implementation-independence***. The reference architecture focuses on functionality and its composition without being tied to specific approaches, technologies or platforms. This makes it possible to compare different implementations.

➢ ***Orthogonality***. The functions integrated in the architecture are independent and do not overlap. This property makes it possible to build customized solutions by removing functions from the general architecture when they are not needed.

➢ ***Completeness***. The architecture includes all the functionality needed to meet all the reasonable requirements of an AN agent. This property can only be validated empirically, by showing that the reference architecture is sufficiently general to encompass existing solutions.

In addition, based on the core characteristics of a telecom network, the agent architecture design must adhere to the following principles:

➢ ***Hierarchical autonomy***. The architecture can be instantiated as an agent at different network hierarchical layers [14] (Resource, Service, Business). Every agent can perform closed-loop operations as needed.

➢ ***Learning from the environment***. The agent can receive information from its environment, such as humans, the network or other agents. It can learn by analyzing its input data, and produce and manage knowledge.

➢ ***Technology agnostic solution*** There are no restrictions on agent implementations, which can range from monolithic AI-based architectures to hybrid architectures integrating model- and data-driven components, looking for trade-offs in reliability and trust.

➢ ***Collaborating with humans***. The agent can interact with humans and collaborate on operational and management tasks. In any situation, humans can take over the agent's functions if necessary.

➢ ***Collaborating with other agents***. The agent is able to interact with other agents, both within the same autonomous domain [15] and between different domains.

## 2.2   Reference architecture

### 2.2.1   Agent interaction environments and flows

We start with a simplified view of the functional reference architecture of the autonomous networks (AN) agent to provide a global overview. The architecture is largely inspired from an architectural characterization of autonomous agents studied in [16, 17, 3]. In this section, we briefly explain the environment with which the agent interacts and how the interaction flow works. Further details on the flow will be provided in subsequent sections.

The AN agent is a system that interacts with three different environments, as shown in Figure 3. (1) The network environment, that is the part of the network it controls, and consists of all the network elements with which it constantly interacts to ensure that its goals are met. (2) The human environment, composed mainly of operation and maintenance (O&M) engineers who interact with the network, issuing directives mainly to deal with problems that may arise during network operation and to adapt to changes in operating conditions. (3) The agent's environment, composed of the other agents with which the agent must cooperate to autonomously achieve the global network goals.

The agent architecture integrates seven functional modules that can cover all aspects of autonomous operation with minimal human intervention. Ideally, the human environment can be eliminated if the tasks of human operators can be systematized and reliably executed by the agent. Among the seven functional modules, the World Knowledge (WK) module plays a special role. It is a knowledge repository containing the knowledge shared and updated by the other modules.

The flow of the agent's interaction with each environment is illustrated in Figure 3. Interaction with the network takes place in a closed loop involving inputs that are first processed by the Situation Awareness module, then by the Decision-Making module, which in turn produces the corresponding outputs. We refer to this closed loop as the agent's reactive behavior. Interaction with the human environment takes place via the Human-Agent Interaction (HAI) module. Finally, interaction with the other agents takes place via the Agent-Agent Interaction (AAI) module. Outputs of the Human-Agent Interaction module are taken into account by the reactive behavior by using two modules, the Self-Awareness and Choice-Making modules, connected in series. They drive the agent's proactive behavior, which generates new goals to be processed by the Decision-Making module.

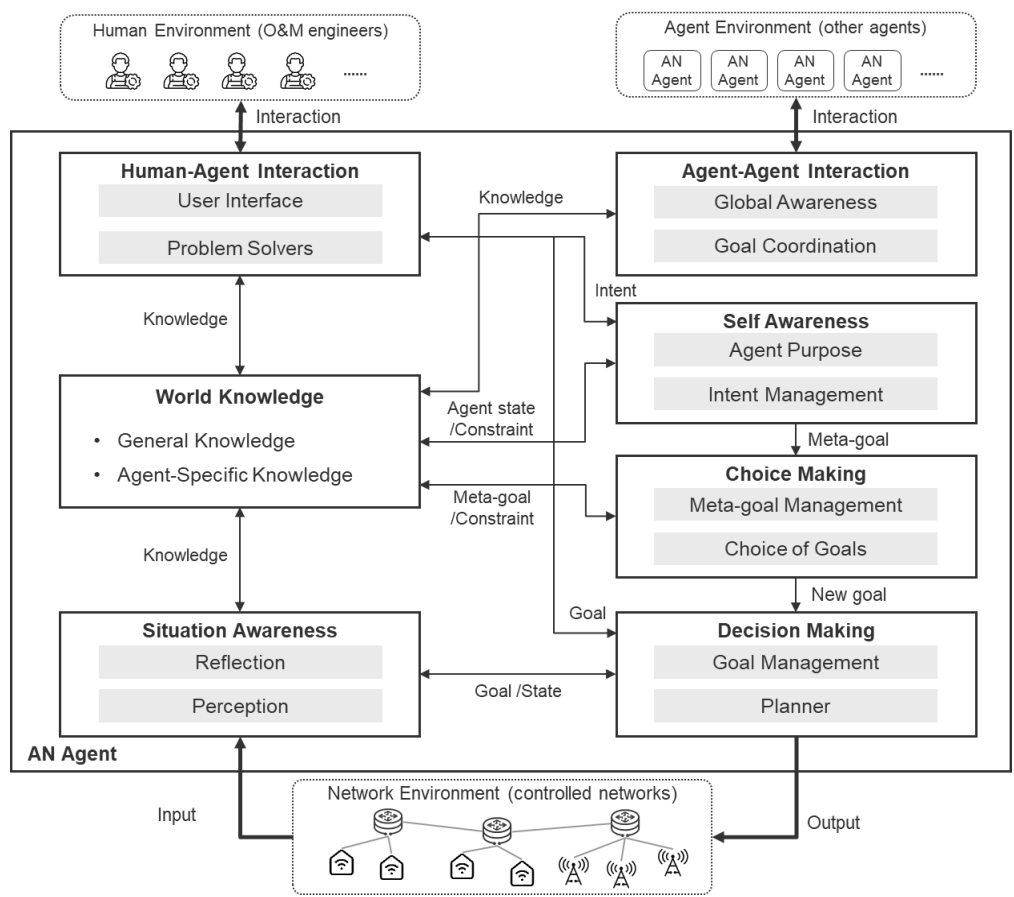

Figure 3 A simplified view of the reference architecture of an autonomous agent

A brief description of the role of each module is given below.

The Situation Awareness module perceives and analyzes information from the network environment to estimate the corresponding network state. The Decision-Making module then determines the appropriate actions to be carried out to achieve the agent's goals, based on the estimated state of the network. Together, these two modules realize the agent's reactive behavior, working synchronously with the network environment to ensure that the agent's goals are achieved.

The Self-Awareness module recognizes potential risks and threats in the controlled network taking into account the agent's state and knowledge. To do this, it selects a set of goals to overcome the detected problem and passes them on to the Choice-Making module. After a cost-benefit analysis, the latter selects the most suitable goal and passes it on to the Decision-Making module for processing.

The roles of reactive and proactive behavior are therefore complementary. The former is designed to ensure real-time operation of the network service. The latter deals with potential risks and threats that could lead to a gap between the service and expectations and is responsible for finding workarounds. The other two modules ensure the agent's harmonious integration into the global network and the human environment. The HAI deals with symbiotic collaboration with humans, and the AAI ensures global awareness and consensual decision-making, enabling collective intelligence.

It should be noted that the behaviors described above all rely on the necessary support of WK. In the following sections, we will analyze each module in detail and explain how, as a whole, they make up an AN agent.

### 2.2.2 Reactive behavior

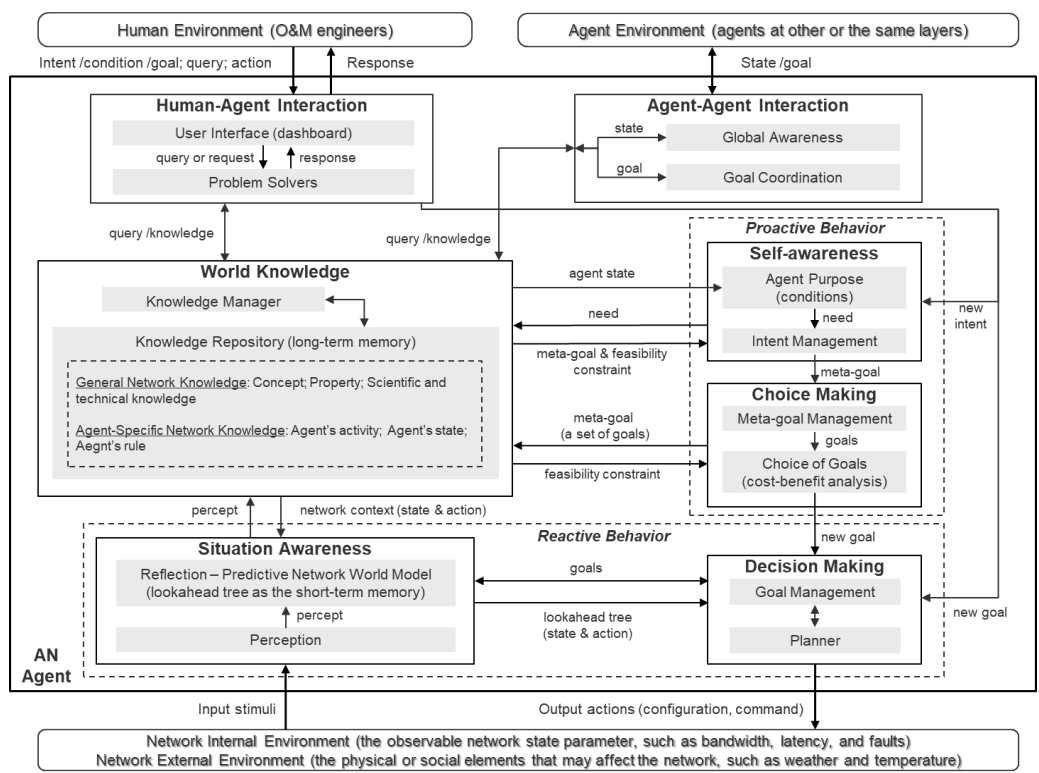

Figure 4 A detailed view of the reference architecture for the autonomous network agent

Figure 4 shows a detailed view of the architecture that refines that of Figure 3. This section explains in detail how reactive behavior processes incoming inputs and produces the corresponding outputs.

The reactive behavior responds to environmental stimuli and is based on two modules: Situational Awareness and Decision-Making.

***The Situational Awareness module*** combines two functions as explained below.

***The Perception function*** first receives stimuli from the network environment as the input, i.e. the observable information from the controlled network. These stimuli consist of (1) information from network devices, such as traffic, KPIs, alarms and logs, or (2) external monitorable information likely to affect the network, such as weather and temperature. The Perception function analyses the stimuli to produce a percept comprising relevant information about the observed state of the environment. Analysis can be performed using machine learning techniques. As explained below, the perceived information can be enriched with knowledge from the World Knowledge module.

***The Reflection function*** receives a percept from the Perception function and first sends it as a query to the World Knowledge module, which responds by generating the most general context corresponding to the percept. The latter contains knowledge that can be used to build a predictive model of the agent's environment, including an estimate of its state and possible interactions between the agent and its environment. The complexity of building a predictive model depends on several factors, including the degree of dynamism of the agent's environment, and we we will not go into that here.

***The Decision-Making module*** combines two functions as explained below.

***The Goal Management*** function manages a given set of goals. It uses the predictive model produced by the Reflection function to decide which goals are applicable based on the state of the model. Note that the set of applicable goals may contain contradictory goals. The Goal management function must therefore determine a maximum set of non-conflicting goals by applying an optimization criterion. For each applicable goal in this set, it activates the Planning function.

***The Planning function*** for each goal transmitted by the Goal Management module, calculates a control policy which generates sequences of actions in response to the actions of the environment, to achieve the desired goal. A control policy is calculated by analysing the predictive model, and may involve both the possibility of achieving desirable conditions and safety constraints involving the avoidance of risky states.

### 2.2.3    Proactive behavior

The proactive behavior is determined by the agent's state conditions and is designed to replace humans in most network operations and maintenance tasks. It would be implemented by two modules: Self-Awareness and Choice-Making.

The Self-Awareness module combines two functions: Agent Purpose and Intent Management.

***The Agent's Purpose function*** controls the degree of satisfaction of a set of conditions that characterize the agent's rationale and depend on the agent's state. Failure to satisfy some of these conditions may compromise the agent's ability to fulfil its role in providing the appropriate service. The agent's state is stored in the World Knowledge module, as explained in Section 2.2.6. Failures of the conditions are critical events that threaten network services, such as increased latency due to aging infrastructure. The Agent's Purpose function for each failure generates a need, the satisfaction of which is managed by the Intent Management function.

For a given need, the ***Intent Management function*** sends a request to the World Knowledge module, which provides a corresponding meta-goal and a set of associated feasibility constraints. A meta-goal is simply a set of goals, each of which is a possible means of satisfying the need. The feasibility constraints include ODD's (operational design domain), normative rules implied by regulations and O&M expertise (see Section 2.2.6). Based on the feasibility analysis of these constraints, the Intent Management function will determine whether the need can be satisfied. If so, a meta-goal specific to this need will be generated and passed on the Choice-Making module.

The Choice-Making module combines two functions: Meta-goal Management and Choice of Goals.

***The Meta-goal Management function*** receives the meta-goal and, in interaction with the World Knowledge module, analyzes the feasibility of concrete goals that can be considered as the realization of the meta-goal. This function returns a set of achievable goals that is passed on to the Choice of Goals function.

***The Choice of Goals function*** takes the set of goals provided from the Meta-goal Management and uses a value system, explained in section 2.2.6, to perform a cost-benefit balance analysis for each goal. This analysis takes into account the actions needed to achieve the goal and the balance of values for the agent and its environment. The choice of a goal may be blocked by the application of normative rules. For example, if the value balance is negative for the environment but positive for the agent, the latter may reject it in order to comply with the rules. Based on the analysis, the most appropriate goal is selected and sent for processing by the Goal Management function.

## 2.2.4   Human-Agent interaction

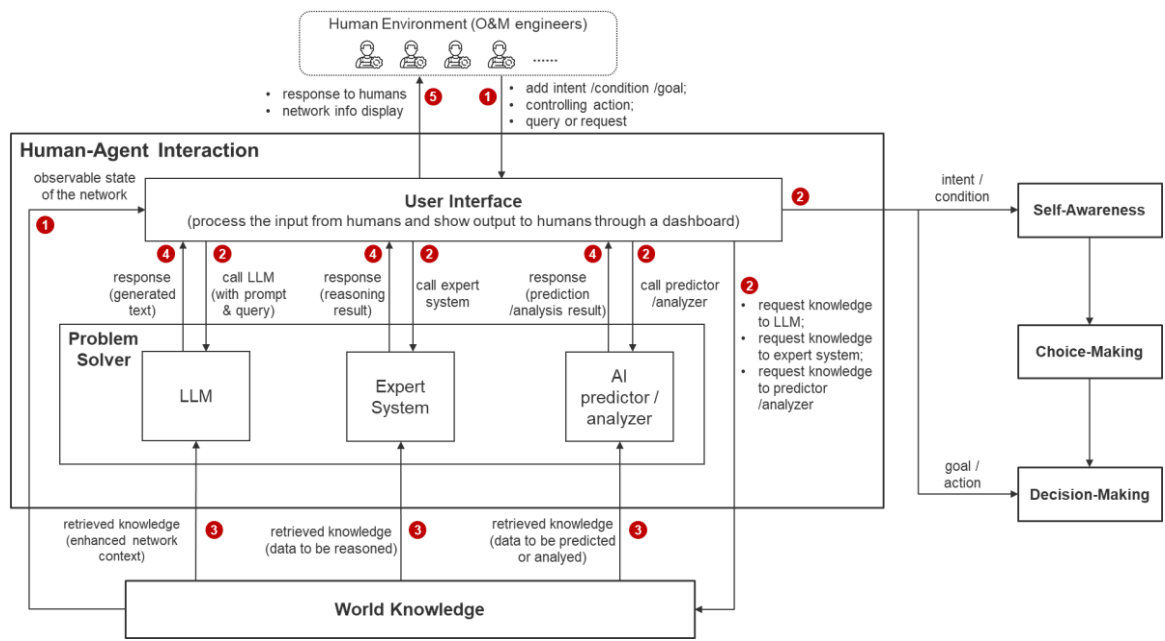

Figure 5 The interaction flow of the Human-Agent Interaction module

The role of humans in the autonomous network is expected to evolve over time. Network O&M tasks have not yet been fully automated and humans remain responsible for network control. During this stage, humans should be equipped with a digital assistant as a copilot to aid in addressing various O&M tasks, such as providing suggestions and analysing data. As human skills become augmented by AN agents (through reactive and proactive behaviors), there will be a reduction of manual interventions and humans will shift focus to the definition of goals, conditions, intents, and strategies. However, humans will retain the authority to take full control of the agent system whenever they deem it to be necessary. Additionally, hardware repair and replacement will continue to require human involvement. To facilitate this symbiotic and progressively changing relationship between humans and agents, a dedicated module is needed in the architecture.

The Human-Agent Interaction module is specifically designed to support the interoperation between humans and agents, which consists of two functions:

***The User Interface*** provides a human-agent interaction language to ensure ease of use. It must support the following functionalities with corresponding inputs and outputs, as shown in Figure 5.

➢ It receives the observable state of the network from the World Knowledge as input and presents synthesized information such as KPIs, alarms, and resource states as the output to humans, through a physical dashboard.

➢ It enables humans to input new intents, conditions, and goals in natural language. These inputs are then recognized and converted into the prescribed format, subsequently being passed to the the Goal Management module and the Intent Management module to trigger reactive and proactive behaviors respectively.

➢ It receives requests from humans as input, and passes them to the Problem Solver after processing. The response generated from the Problem Solver is returned to the User Interface and shown to humans through the dashboard.

➢ It allows humans to directly input specific actions for network control, such as configuration commands. These actions are passed to the Decision-Making module for execution.

➢ It also allows humans to provide feedback to the agent. This feedback is passed to the agent's Self-Awareness module so that the agent can learn from this feedback.

***The Problem Solver*** module is a "digital network expert", receiving inputs (queries or requests) from engineers through the User Interface and invoking the corresponding internal components to generate results based on the input. It consists of three main components: the LLM (Large Language Model) playing the role of a copilot, the expert system, and the specialized AI predictor/analyzer. Each component has distinct strengths and should be deployed in parallel. The choice of which component to invoke depends on the input from humans. The selection of the appropriate component to generate the response depends on the nature of the input provided by the user.

➢ *The large language model (LLM).* We need a specialized co-pilot for the telecommunication network to optimize decisions and improve efficiency. The LLM is a suitable technique for this task. As known, the LLM excels in capturing the semantic relationship from the data and generating human-understandable answers, so that it can communicate with humans in natural languages. For the LLM specialized in network management, the training dataset must include O&M knowledge to ensure that the model can comprehend various O&M tasks and generate appropriate results. The tasks include such as modifying or improving configuration command scripts, generating reports, recommending packages, and offering optimization suggestions. Moreover, the LLM needs to be trained to understand the knowledge about the network context such as actions, goals, and states, from the World Knowledge by using the RAG architecture. By doing this, it can generate practical solutions considering real-time network context and goal achievement, rather than simple question and answering. For instance, engineers can ask "I want to reduce the latency of region A by 2% but without affecting other KPIs under the current network context, provide concrete configuration commands to me".

➢ *The Expert System (ES)*. The ES can represent prior domain knowledge from experts as particular rules, and then process reasoning based on the rules. It is essential to codify the extensive experience accumulated by O&M experts into formalized rules to enhance the efficiency of other O&M personnel.

➢ *AI Predictors and Analyzers*. The decoder-only LLM excels in generative tasks but may not be sufficient for tasks requiring deep understanding. Many O&M tasks require a deep understanding of large volumes of input data, followed by the generation of simple but precise outputs, rather than the production of extensive content. These tasks can be classified into two types: (1) prediction tasks, such as traffic forecasting and port failure probability prediction, and (2) analysis tasks, including root cause analysis of alarms and device log analysis. Therefore, we need an AI-based analyzer and predictor. While the number of parameters need not be large, the model's architecture must be specifically designed to accommodate the structure of the input and output data. It must fully comprehend the input information and generate highly accurate outputs.

### 2.2.5 Agent-Agent interaction

A communication network is a typical distributed system, where devices and management systems are deployed according to specific topologies, either physically or logically connected, as shown in Figure 6.

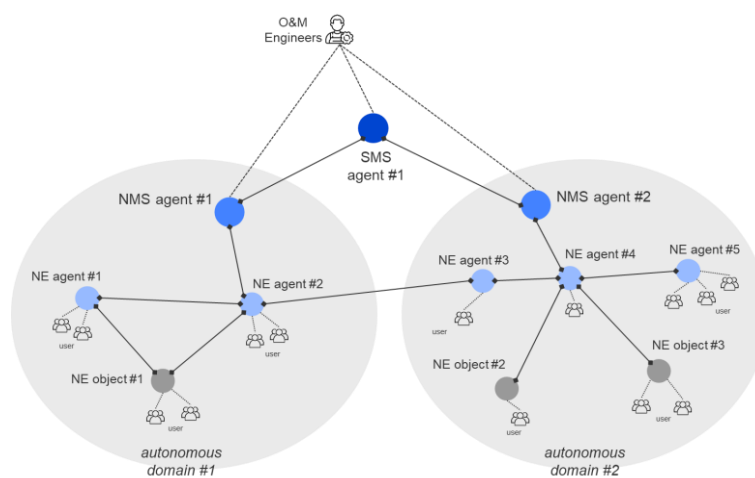

Figure 6 An example of that topology that different types of AN agents connected

In a network topology, multiple network elements (NEs) are typically present. A specific network management system (NMS), oversees the management of these NEs. A single NMS, along with the associated NEs, can constitute one form of the autonomous domain. To manage the services across multiple domains, a higher-level network management system is employed. This system, which manages several instances of NMS, is termed the service management system (SMS). The system's naming rules refer to the definition in the ITU-T standard M.3010 [18].

In AN, these systems will become agents. Most network service delivery and operations and maintenance (O&M) tasks are accomplished by multiple agents through their interactions, as a single agent cannot handle these tasks independently.

To meet these requirements, a specific module is needed in the agent architecture. The Agent-Agent Interaction module is specified to achieve network collective intelligence, which consists of two functions: Global Awareness and Goal Coordination, as shown in Figure 7.

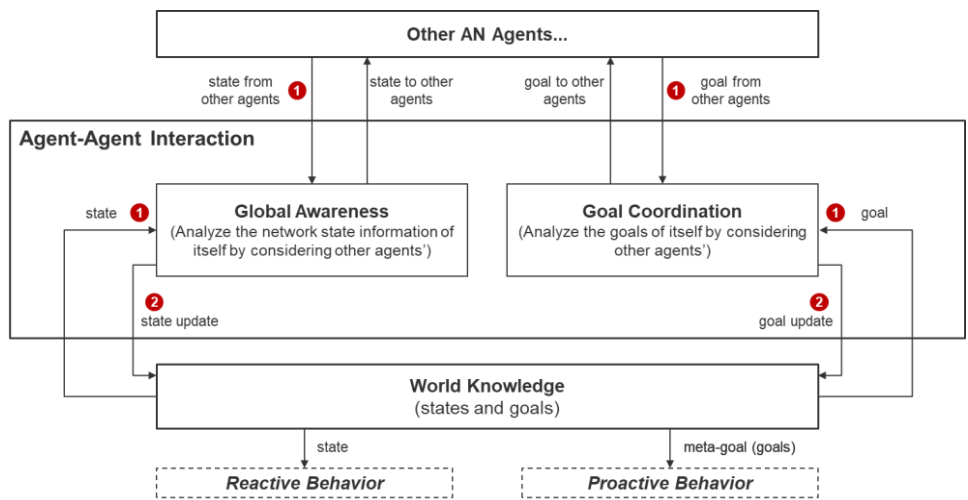

Figure 7 The interaction flow of the Agent-Agent Interaction module

***The Global Awareness function*** enables an agent to understand situations of other agents. Specifically, agents can proactively exchange state information through interaction. The relevant state information to be exchanged depends on the specific situations encountered. The function performs a comprehensive analysis by integrating the state from other agents into the agent's own state. Its output is then used to update the corresponding state in the World Knowledge module. The state update will affect the follow-up reactive and proactive behavior

***The Goal Coordination function*** can receive goals from other agents at the same or higher operational layer (as discussed previously the autonomous network architecture is stratified into three layers - a business operations layer, a service operations layer, and a resource operations layer) and send its own maintained goals to others. The purpose of this function is to analyze the relationships between their own goals and those of other same-layer agents, especially on the two aspects:

➢ *Conflict*: The function identifies potential conflicts between the goals of the various agents. A conflict refers to the pursuit of a goal that hinders the achievement of another goal, which is common in complex networks.

➢ *Synergy*: The function identifies mutually beneficial goals between agents, where the achievement of one agent's goal facilitates the achievement of another agent's goal.

The result of the analysis determines whether it is necessary to update the existing goals in the World

Knowledge module. The goal update will affect the follow-up reactive and proactive behavior.

## 2.2.6    World Knowledge

The term "knowledge" refers to the information that enables the network agent to interpret stimuli in its environment and adapt its decision-making processes to changes in it. The **_World Knowledge_** (WK) module is used to store and manage knowledge, and consists of two parts: (1) the Knowledge Repository, used to store all types of general and agent-specific knowledge; and (2) the Knowledge Manager, which is a function for managing knowledge in the repository and interacting with other modules, as explained below.

The knowledge contained in the **_Knowledge Repository_** can be grouped into two categories:

_The General knowledge_, which is common to all network agents and includes concrete and symbolic data of various types, such as,

➢ Taxonomies containing domain-specific terms about networks, their types, topology and software and hardware components;

➢ Properties that agents must satisfy, such as technical requirements concerning their safety, security and performance;

➢ Scientific and technical knowledge, including rules derived from standards and libraries of problem-solving methods.

_Agent-specific knowledge_, relevant to the agent's role, such as

➢ _The specification_ of possible agent activities that can be used by the agent to achieve its goals: (1) a list of possible actions with their enabling conditions; (2) a list of possible goals, (3) a list of needs with the corresponding meta-goals for achieving them.

➢ _The agent state_, which includes relevant information on: (1) the state of resources available in the network controlled by the agent; (2) the list of goals pursued, together with the list of goals managed by the Goal manager; (3) the values of KPIs calculated by the network's lower layers.

➢ Specifications of the _rules_ guiding the agent's decisions and choices, and in particular (1) the operational design domain (ODD), which defines the operational scope and limits of the agent's behavior; (2) the normative rules, which define the legal and ethical rules that should guide the agent's choices; (3) and knowledge summarizing the engineer's experience, which defines rules of thumb that the agent can use to cope with critical situations.

➢ _A value system_, i.e., a set of value scales and rules that the agent can use to estimate the costs and benefits of its actions in terms of economic or immaterial values. Value systems express different types of constraints imposed by the agent's environment. Costs can be expressed in monetary terms, or in terms of reputation and trust on the part of other agents. For example, to create synergies with other agents, it is important to behave ethically. A selfish agent will have to face the distrust of its peers, because its choice brings it a certain benefit while entailing excessive costs for other agents and the environment. The implementation and effective use of a value system similar to value systems used by humans raises technical issues such as establishing correspondences between value scales and the representation of value-based decision rules [19] [20].

**The Knowledge Manager** function manages interaction between WK and other modules. It consists of three parts:

➢ *The retrieval mechanisms* that handle the different types of input queries from other architectural modules, and provide the corresponding output as the retrieval knowledge, as shown in the Table 1 below.

Table 1 Different input and output for the World Knowledge module

| Interacts with | Input | Output |
|---|---|---|
| Situational Awareness | percept | network context (perceived states and actions) |
| Self-Awareness | need | feasibility constraints (agent's rules) |
| Choice-Making | meta-goal (context) | feasibility constraints (agent's value system) |
| Human-Agent Interaction | natural language | embedding vector |
| Agent-Agent Interaction | request | relevant state information and goal |

➢ *The update mechanism* that based on the input query keeps knowledge in WK up-to-date and valid. In particular, it is important to update the agent state from percepts and interaction with human and agent environment.

➢ *The coherency-checking mechanism* integrates new knowledge in the repository and ensures that World Knowledge remains logically coherent and consistent.

## 2.3 Hierarchical network architecture

### 2.3.1 Network hierarchy

Referring to the layer division principle in the network standard organization [15], the hierarchy from top to bottom can be divided into three interrelated layers (as shown in Figure 8), to support the network operation and service delivery: (1) The Business Layer, which defines the company's strategic objectives and service requirements, and guides network design and maintenance operations; (2) The Service Layer, which provides specific network services based on the company's requirements in several autonomous domains; (3) The Resource Layer, which provides the domain's physical and virtual infrastructures to deliver the service.

In autonomous networks (AN), network systems distributed at different layers in the network hierarchy should become agents. The AN agent reference architecture should serve as a general framework for the instantiation of these agents.

In what follows, we will explain the devices and systems present in the layers of the hierarchy, along with the scope of the network they control.

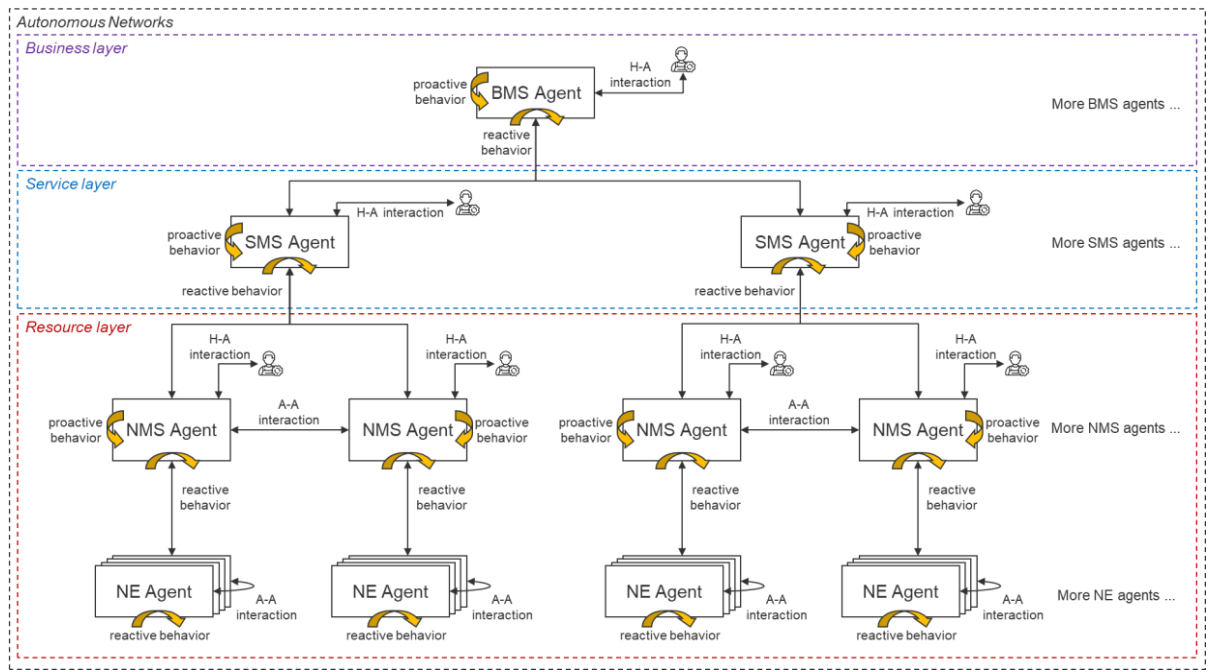

Figure 8 The overall network hierarchy with agents

***Business layer***. An agent instance at the Business layer is typically referred to as the Business Management System (BMS) agent. This is the highest-layer agent, connected to multiple Service Management System (SMS) agents, with its controlled network encompassing all other agents across the Service and Resource layers.

***Service layer***. An agent instance at the Service layer is known as an SMS agent. It controls several autonomous domains and is therefore connected to several Network Management System (NMS) agents under its control.

***Resource layer***. Both network element (NE) and NMS agents are located at the Resource layer. The controlled network of the NMS agent consists of multiple NEs agents. The NE controls its components, such as ports, main boards, memory, and links. In this layer, two types of most common agent-to-agent interactions are: (1) between NE agents within the same domain and (2) between NMS agents managed by the same SMS agent.

With the exception of the NE agent, other agents can be capable of interacting with operation and maintenance engineers via the Human-Agent Interaction.

Within the hierarchy, agents at adjacent layers can interact with one another for information exchange. These agents collectively form an overarching architecture across layers, called hierarchical architecture, as illustrated in Figure 9.

The following sections describe how the reference architecture can be instantiated as different agent instances within the hierarchical architecture. The primary focus of our research and discussion is on the NE and NMS agent instances.

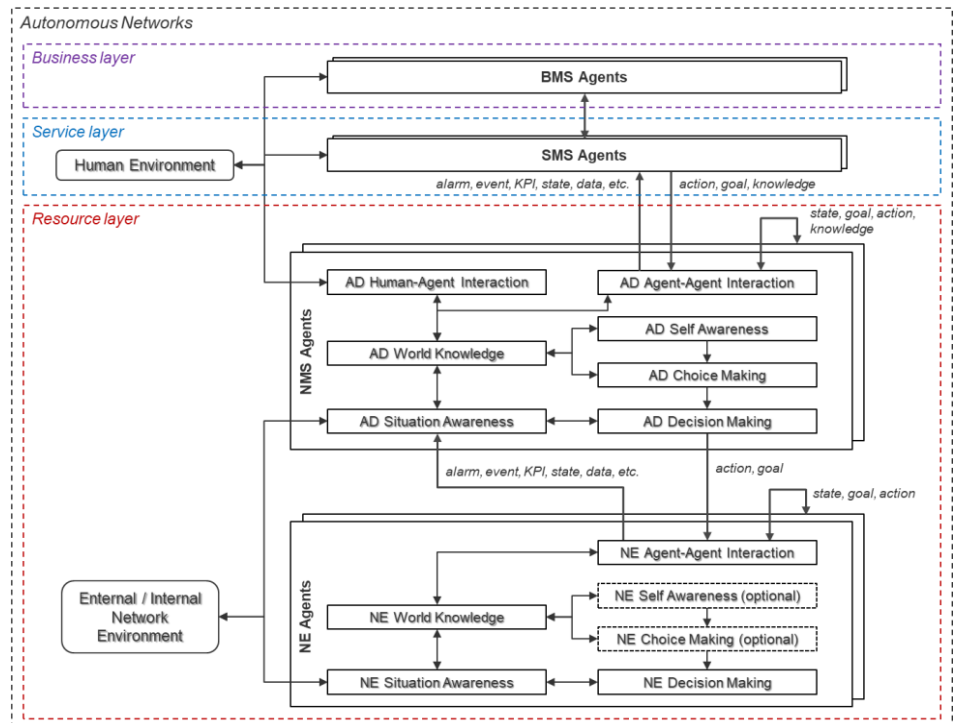

Figure 9 Hierarchical architecture consisting of multiple agent reference architectures

### 2.3.2 Agent instance for the NE

Reactive behavior (RB) is essential for the NE agent, encompassing functions that support the operation of network services, such as routing and forwarding, as well as mechanisms that ensure service continuity, such as switchover and congestion mitigation. RB enables the NE to maintain its operational goals within the scope of its controlled network. In contrast, Proactive behavior (PB) is optional for the NE agent. This is mainly due to two considerations: firstly, the execution of proactive behavior demands additional knowledge and computational resources, which are constrained in the NE; secondly, the PB can be delegated to the NMS agent.

The instantiation of interaction modules is contingent upon the specific design of the inter-agent interactions. In the AN, all NEs are managed by the NMS agent. Even when human intervention is required for accessing a specific NE, this is facilitated through the Human-Agent Interaction (HAI) module of a NMS, rather than direct interaction with the NE itself. The HAI module can therefore be omitted. However, frequent interactions between NEs are essential to the realization of collective intelligence, achieved by an Agent-Agent Interaction (AAI) module.

The NE agent knowledge is local. Each NE agent retains its own knowledge, which is generally not shared or synchronized with other agents, unless explicitly requested. Even within the same domain, the knowledge of different NE agents can vary.

### 2.3.3 Agent instance for the NMS

RB and PB are integral parts of the NMS agent. RB performs operation and maintenance (O&M) tasks with predefined goals, enabling immediate responses to the controlled network to ensure service continuity. These include tasks such as path finding and the joint optimization of several parameters. PB, on the other hand, enables the NMS agent to proactively manage and maintain the network, dealing with potential risks that could disrupt or degrade the network before damage occurs. PB requires additional knowledge and involves the creation of new goals.

The HAI is necessary for the NMS agent It enables the engineer to provide on-demand assistance, feedback and facilitates efficient management of goals and conditions by humans. Moreover, the AAI allows the NMS to exchange information between different NMSs, thereby being aware of states and goals from other agents. The perception, goal, and action information exchanged between domains is substantially more complex than that shared between individual Network Element (NE) agents. For instance, decision-making information transmitted from one NMS agent to another may involve a sequence of actions that engage multiple controlled NE agents.

As far as world knowledge is concerned, the NMS agent's World Knowledge must encompass the entire operational domain, with particular emphasis on specific areas such as the agent state. The agent state reflects the overall condition of the controlled network within the domain, as well as the ongoing activities related to its goals, actions, and operational requirements. Additionally, it includes the operational rules governing the domain. This comprehensive body of knowledge enables the NMS agent to effectively manage and optimize the network it controls.

### 2.3.4   Agent instance for the SMS and BMS

The Service Management System (SMS) and Business Management System (BMS) agents should also follow the hierarchical connection rule. They are deployed at the Service and Business layers respectively, and their implementation should also follow the reference architecture pattern in principle.

Different from NE and NMS, they are oriented towards end-users and business objectives. Due to the distinct role of their respective layers, their functional characteristics may diverge from those of NMS and NE agents. A comprehensive analysis of the agents at the Business layer and Service layer will be carried out in future research.

## 2.4   Distribution of roles between reactive, proactive and human-controlled behavior

### 2.4.1   Behavior role distribution principles

Given the high complexity and large scale of modern networks, network operation and maintenance tasks are highly diverse, with most of them currently being performed by human operators. When implementing our proposed architecture to achieve real-world network O&M tasks, a crucial question arises: What tasks should be handled by the reactive behavior (RB), proactive behavior (PB), or Human-Agent Interaction (HAI)?

A nominal event refers to a situation where the system is operating under normal, expected, and designed conditions while an off-nominal event occurs when the system operates outside its normal or expected parameters. From this point of view, RB typically addresses nominal events that must be processed in near real-time to prevent service disruption. In the context of the self-driving vehicle RB is responsible for achieving a predefined set of goals that characterize driving capabilities, such as maintaining trajectory, making turns, and avoiding pedestrians. On the other hand, PB handles off-nominal events, which occur sporadically and have the potential to interrupt the service. For self-driving vehicle, such events could include a flat tire or the sudden failure of a device, for which predefined safe procedures must be implemented. Those off-nominal events not covered by PB should be addressed by humans. For instance, a low battery or malfunctioning lights necessitate human intervention. In conclusion, the frequency of event occurrences is a key factor in differentiating the responsibilities of RB and PB. Additionally, RB demands a faster response time, typically near real-time, whereas the response time of PB depends on the nature of the event.

Some events occur sporadically, such as fiber failures or power outages. While they may be classified under PB based solely on their frequency, these events typically lead to significant service degradation or even interruption. From the perspective of ensuring network service continuity, such events must be addressed immediately, aligning more with the responsibilities of RB. Therefore, we introduce an additional dimension: the urgency of the event.

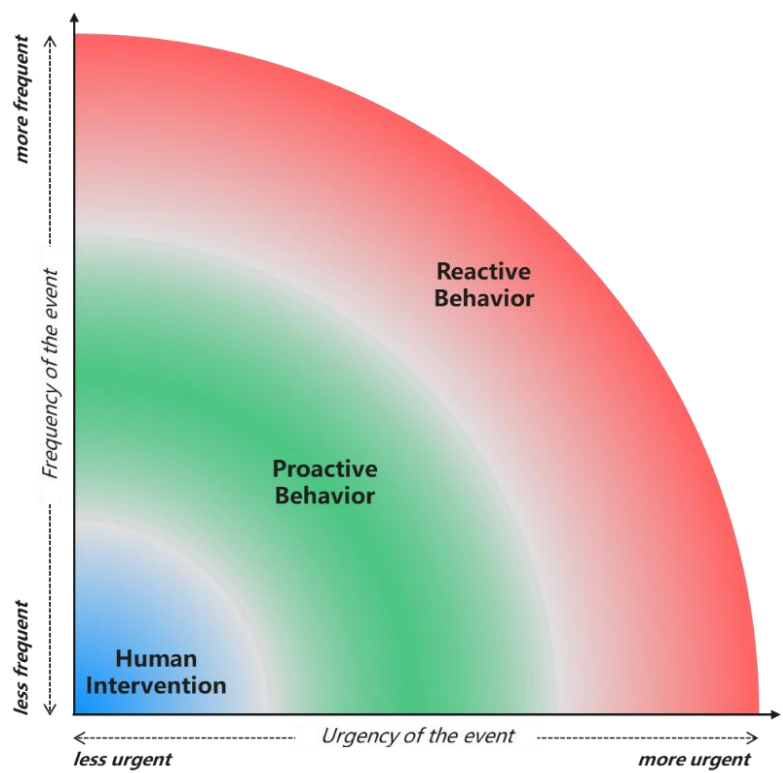

Figure 10 The principle of differentiating the responsibility of behavior by frequency and urgency

We propose a clear principle that delineates the boundaries of network O&M tasks and further distinguishes which tasks should be handled by each behavior. Two dimensions are defined: ***frequency and urgency*** of the events, as shown in Figure 10. In an autonomous network, events with higher frequency or urgency are intended to be automated by RB in the AN agent (the red area). From the perspective of network services, these events are likely to cause service degradation or interruption, such as message loss or a port down. Conversely, events with lower frequency and urgency are addressed by PB (the green area). These events may not immediately harm network services but pose potential risks, such as fiber aging. The management of other events, of lesser frequency and urgency or which the PB cannot cover, must be assigned to man (blue zone). It should be noted that frequency and urgency are not positively correlated with importance, less frequent and urgent tasks can also be important.

Finally, it is important to emphasize the hierarchical relationship between the three types of behavior. At the highest level, HAI behavior can generate intents and goals that will be processed by PB and RB, respectively. In turn, PB can generate goals for RB.

In real-world networks, network designers and O&M engineers must carefully define clear boundaries based on frequency and urgency to distinguish events handled by RB, PB, or humans. To note, these boundaries are not absolute; rather, they should be flexible to a certain extent, allowing for gradual shifts as the network operates. We illustrate two cases: (1) An O&M task done by PB may shift to RB depending on the network conditions. For instance, when a network is deployed, the infrastructure is new, and the number of services being carried is relatively low, resulting in a minimal frequency of failures. In such cases, service restoration following a failure can be handled by PB. However, with hardware components aging and the number of services increasing, the frequency and urgency of failures may also rise. Consequently, the responsibility for service restoration may shift to RB to meet the growing demand for fast response and recovery. (2) The same task may be handled by different behaviors in different networks, which also depend on the actual network conditions. For example, if the network infrastructure in domain A is newer and of higher quality than in domain B, with a much lower probability of failure, service restoration may be addressed by PB in domain A and by RB in domain B.

### 2.4.2    Behavior role distribution in network use cases

Based on the proposed principle, we give a practical network use case as an example to illustrate how RB and PB are applied in the AN agent architecture.

The use case is about private line service in the optical network. The private line service refers to a dedicated and, high-bandwidth communication service running on optical network elements (NE). Service messages are transmitted from the source and are ultimately received by the sink NE through predefined route paths. We examine how the behavior of AN agent reference architecture can be applied to implement the autonomous O&M in the use case and discuss the role of different behaviors based on the proposed principle.

#### *The role of Reactive Behavior*

The "1+1" service is a type of private line service, where the line is configured with two routes: a working route and a protection route, represented by orange and purple curves in Figure 11, respectively. Messages are transmitted simultaneously over both routes. Under normal conditions, the sink NE receives messages from the working route. The RB of the NE handles message sending and receiving due to the high frequency of communication. If, for example, the port on NE#3 fails, causing the working route to be unavailable, the sink NE must immediately activate the switch-over mechanism to receive messages from the protection route, thus avoiding service interruption. This switch-over is the responsibility of the RB of the NE, as it requires the highest urgency.

However, the service is now in a degraded state as there is only one available route for the private line and it is not protected anymore: the service will be interrupted if this route fails. The NMS agent must quickly find an available new protection route for the private line, and this is the role of the RB of the NMS agent due to the urgency of the task. This process can be broken down as follows. Situation Awareness first gathers relevant information from the controlled network, which consists of multiple NEs. In this use case, the information typically includes port states, board states, fiber statistics, message correctness, and more. The Perception function generates a percept, identifying that the port failure has caused the working route of the private line to fail, leading to service degradation. This percept is used as a query to search for the corresponding network context in the World Knowledge. The context indicates that the current state of the private line does not meet the "1+1" requirements, repairment is to be done. Next, the Reflection function constructs a predictive model based on the context (network states and possible actions). Situation Awareness collaborates with Decision-Making to activate those applicable goals (within Goal Management). For instance, goals can be to delete the old working route and create a new protection route. Based on the predictive model, the Action Planning function calculates a policy. In this use case, this policy must be capable of selecting an available route that satisfies the requirements for the private line. The policy then generates a sequence of actions to configure the route, which involves creating or deleting cross-connections between the internal logical ports of the NEs.

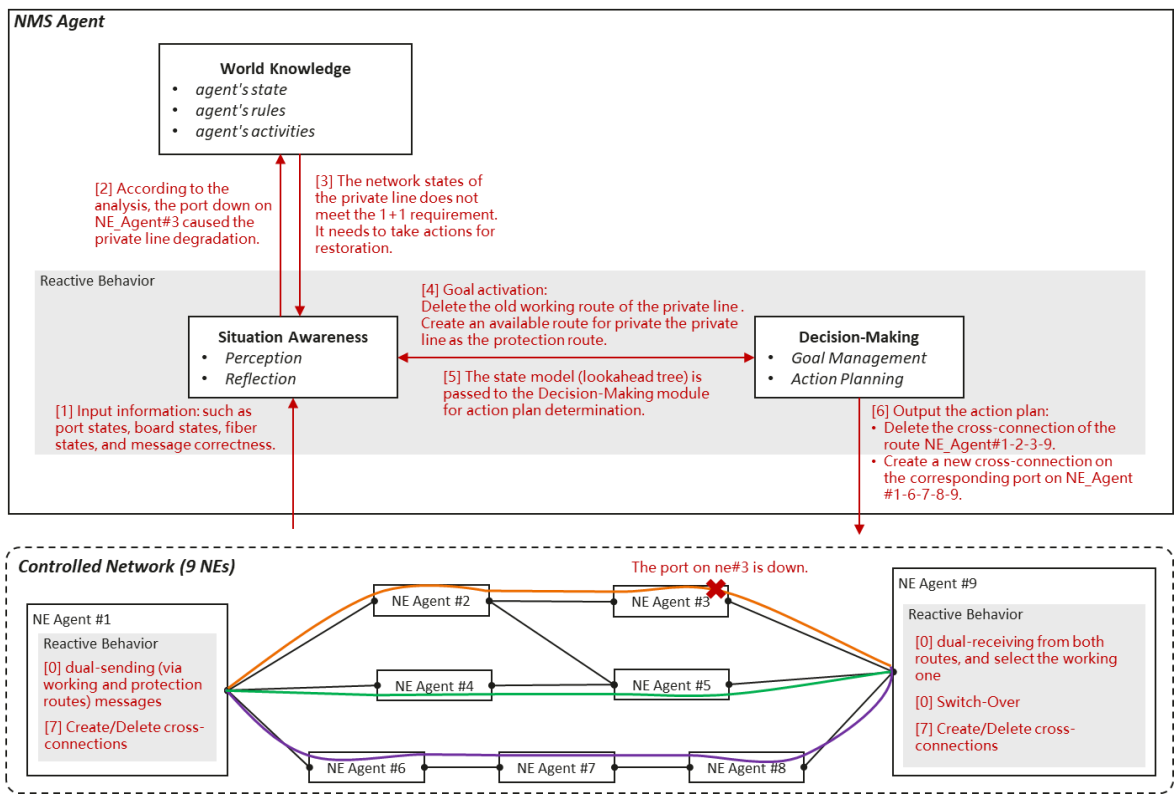

Figure 11 The illustration of applying Reactive Behavior into optical network private line use case

Afterward, corresponding NE agents receive the actions from the NMS agent. The NE's RB instantly deletes the old cross-connections of the old working route and builds cross-connections of the new protection route.

In general, the RB of NEs such as switch-over, cross-connection deletion, and creation, must be done in milliseconds, to maximize the service continuity and avoid any interruption. the RB of NMS, such as deleting the unavailable route and finding a new route should be complete in seconds, which minimizes the duration of service degradation. In the autonomous network, we expect that the whole process is done by RB of NE and NMS agents instantly, the fault is not perceived by private line users.

### *The role of Proactive Behavior*

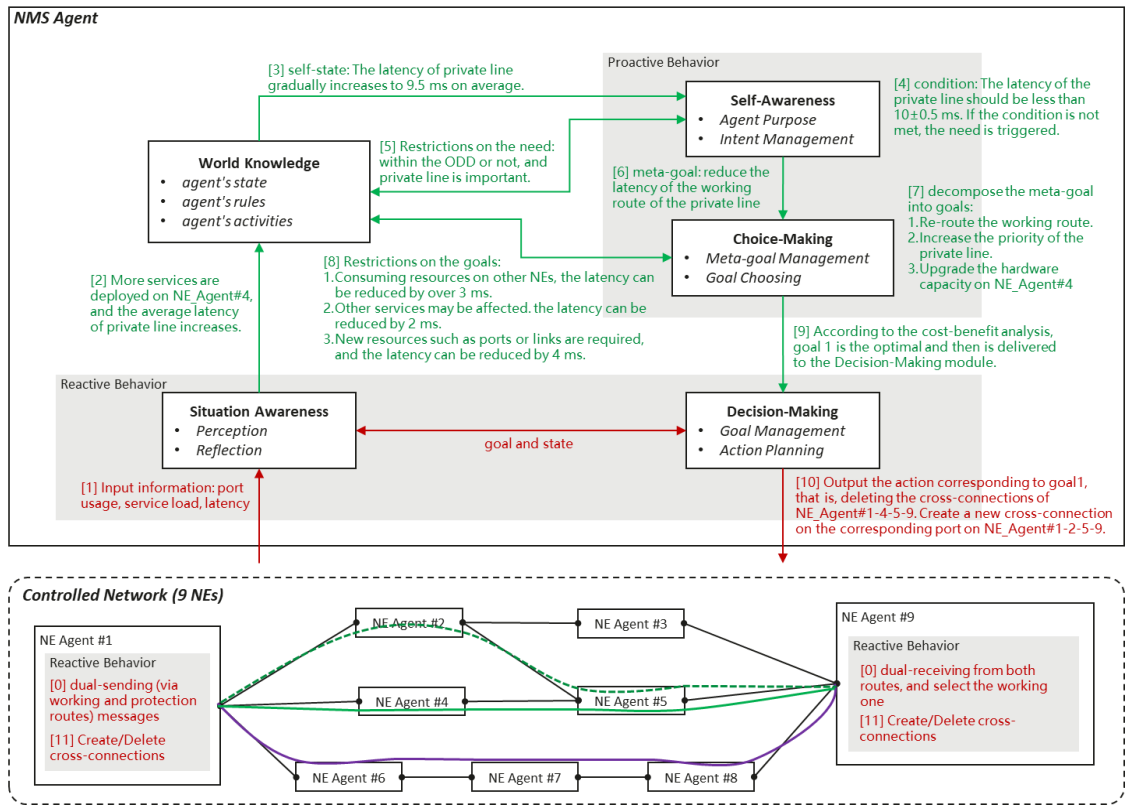

Figure 12 The illustration of applying Proactive Behavior into optical network private line use case

As the network infrastructure ages and the number of services increase, the network load grows. This leads to a gradual decline in performance and an elevated risk of service degradation. Continuing with the private line use case, the current working route is NE#1-4-5-9 (the green solid curve in Figure 12). If the service on NE#4 increases and the hardware components age, the average latency of this route could rise from 6 ms to 9.5 ms. Assume that one of the customers' requirements is that the latency must be less than 10ms with high availability. At this point, the service continues to function and still meets the requirements. However, there is a potential risk of deterioration, degradation, or interruption in the future. The agent controlling the network must proactively improve the system to mitigate this risk. This responsibility falls to PB, as the situation is comparatively not urgent and not frequent. We break down the process of PB as follows.

The information is received by the NMS agent and used to update the knowledge in the World Knowledge (WK), where the knowledge represents the agent state, providing a comprehensive understanding of the entire controlled network. The Self-Awareness module continuously monitors the agent state within the WK. In the agent's Purpose, assuming a predefined condition for the private line service (e.g., latency should remain <10 ms), an increasing trend in latency is detected in the agent state, which will cause the service to fail to meet this condition. This triggers the need to address the latency issue.

Next, the Intent Management function sends the need as a query to retrieve the corresponding knowledge from the WK. This knowledge includes the meta-goal aimed at fulfilling the need, along with feasibility constraints. For this use case, the meta-goal might be, for instance, "reducing the latency of the working route of the private line." The constraints describe factors influencing the achievement of the meta-goal, such as whether reducing the latency is within the operational design domain (ODD), the importance of the private line service, legal and ethical considerations, and prior experience with similar situations. The meta-goal will only be processed if all constraints are satisfied.

The meta-goal can be decomposed into several sub-goals. For instance, Goal 1 is to create an alternative working route to replace the current one, Goal 2 is to increase the priority of this line across NE devices, and Goal 3 is to upgrade the hardware capacity on NE#4. There may, of course, be other potential goals. The Goal Choosing function treats each goal as a query to retrieve the relevant knowledge. The retrieved knowledge includes the feasibility constraints and the value of achieving the goal, as derived from the value system in the WK.

For example, Goal 1 is highly feasible as it only requires utilizing idle resources on other NEs, and it is estimated to reduce latency by over 3 ms. Goal 3, on the other hand, has lower feasibility, as upgrading hardware typically requires human intervention, although it would have the most significant impact on reducing latency. The Goal Choosing function performs a comprehensive cost-benefit analysis, weighing both the feasibility and value of each goal. Due to its relatively high feasibility and value, Goal 1 is selected as the optimal goal and forwarded to the Decision-Making module. In extreme cases, such as when the entire controlled network has very few idle resources available, no goal may be feasible or capable of generating effective value. In such instances, the agent may conclude that the meta-goal cannot be achieved, and human intervention may be required.

This concludes the PB process. Determining the appropriate actions to achieve the selected goal falls under the responsibility of RB. In the following RB, the new route across NE#1-2-5-9 (the green dashed curve in Figure 12) is created, which releases the service load on NE#4, and meanwhile reduces the latency.

In conclusion, the PB of NMS can autonomously optimize the controlled network by stepwise analysis, in order to eliminate the potential risk of affecting the network service.

## 2.5 Incremental evolution of the architecture

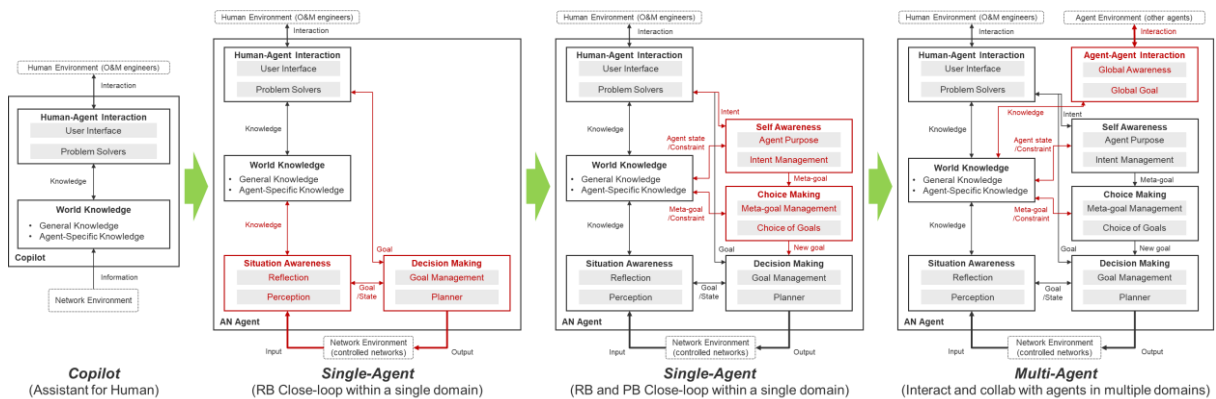

Figure 13 Incrementality of the AN agent reference architecture

The realization of the proposed AN agent reference architecture can be seen as the ultimate goal for a fully autonomous network. However, achieving this is a long-term process, and therefore, the architecture modules should be implemented incrementally rather than all at once. As each module is implemented, the agent's capabilities will gradually improve, ultimately reaching full autonomy once all modules are in place. We propose an incremental achievement development path for this architecture, as illustrated in Figure 12, and the incrementality part of each phase is marked in red. This path can be divided into three progressive phases: *"Copilot", "Single-Agent", and "Multi-Agent"*.

***"Copilot" phase***. This phase represents the most simplified form of the AN agent, consisting solely of the Human-Agent Interaction (HAI) module and the World Knowledge (WK) module. The HAI module is for interaction with humans, providing appropriate responses based on their queries, thus assisting in network operation and maintenance (O&M), as detailed in Section 2.2.4. The WK is the knowledge base, offering knowledge to enhance the quality of responses. This knowledge is continuously updated by gathering information from the network environment. In this phase, network O&M tasks are still handled by human operators, but the efficiency is highly improved with the Copilot.

***"Single-Agent" phase***. The reactive behavior (RB), including Situation Awareness and Decision-Making, is added to the agent. The agent is capable of performing closed-loop control, meaning it can take actions in response to stimuli from the controlled network to achieve its goals without human intervention. The closed-loop is within a single autonomous domain, as this phase is called "Single-Agent". For this, the knowledge needed to support RB is incremented into the WK, including network states, actions, and constraints related to these actions.

After RB, proactive behavior (PB), including Self-Awareness and Choice Making, is further added to the agent. This allows the agent to self-improve by continuously monitoring its agent state and generating new goals, thus enhancing its control capabilities within the domain. Similarly, additional knowledge to support PB should be added to the WK, including the agent state, feasibility constraints and the value system. In this phase, the AN agent replaces humans in handling network O&M task.

***"Multi-Agent" phase***. Finally, the Agent-Agent Interaction (AAI) module is added to the architecture. In this phase, the agent not only controls the network within its own domain, but also interacts and collaborates with agents at the same levels in different domains, exchanging relevant knowledge about states, actions and goals.

# 3 Technical Challenges Ahead

This section discusses the technical challenges of developing the proposed autonomous network reference architecture and presents work directions and their underlying technical challenges.

## 3.1 Achieving trustworthiness

The network operation and maintenance (O&M) is a highly critical, and intentional or unintentional errors caused by humans may result in substantial losses. The AN agent is designed to replace humans and take full responsibility for O&M decision-making. This shift raises a problem: how can we trust the system to make decisions and control the network by itself, i.e., how do we ensure its trustworthiness? The problem can be considered from two aspects of the system life cycle, the design time and the run time.

### 3.1.1 Design time assurance (DTA)

DTA follows the principles of traditional model-based system development, which starts with the analysis of system requirements and progressively determines the system architecture and its components in a top-down flow. This process is generally supported by modeling and verification tools that enable design choices to be justified and their impact assessed. It is often illustrated by the V-model recommended by standards for mission-critical systems [21].

The traditional development approach cannot be applied to ML systems, as they are not explicable and are currently developed on a rather ad hoc basis. System requirements cannot be treated as properties. They are represented by the data used to train the AI system in such a way as to mimic the system's implicit behavior.

It follows that DTA is only applicable to software and hardware systems that can be explained. In our agent architecture, it can in principle be applied to decision-making, both goal management and planning. Although this raises some technical challenges, DTA for these functions is an important step in making an agent's reactive behavior reliable. We can also envisage applying DTA to other model-based components of agent architecture, such as the cost-benefit analysis of the decision-making module and the software used for tracking and knowledge management.

Applying DTA requires a set of domain-specific methods and tools, as explained below.

***Methods and tools for specifying and analyzing network properties***. Networks are distributed systems with a hierarchical structure, for which we need suitable specification tools that enable us to express requirements and desired properties in a rigorous way. In particular, we need tools based on a network ontology for specifying basic concepts and their relationships, as well as tools for checking the consistency and completeness of requirements.

We identify the following work directions: (1) Specification and validation of network services, characterized by Quality of Service (QoS) properties such as constraints on latency and packet loss. (2) Specification and validation of network infrastructure properties, in particular to express connectivity properties and invariants that must be satisfied by reconfiguration scenarios. (3) Validation of network functions, such as routing functions, switching mechanisms, service provisioning and optimization functions. (4) Specification of network types and use cases, for four network types - IP, wireless, optical and core - and their main characteristics and properties.

***Methods and tools for modeling and verifying network behavior***. Behavioral modeling and validation are often closely linked. For modeling, we need a platform that integrates a domain-specific language (DSL), libraries of component types and links, as well as simulation and validation tools. In particular, the DSL must enable component-based description of network behavior and support network dynamism and reconfiguration. The DSL must also support knowledge-based behavior modeling, which is essential for the description of adaptive behavior. As far as validation is concerned, we need a variety of tools, including verification tools using enumerative or symbolic techniques, performance analysis tools, as well as tools for testing virtual or real network prototypes.

These tools can only be proprietary, as commercially available tools suffer from scalability limitations and do not cover the aforementioned needs. In fact, major technology companies have developed their own modeling and validation technology as part of their strategic assets.

### 3.1.2    Run time assurance (RTA)

AI components will increasingly be used to enable smarter, more efficient automation of network operations. AI breaks with traditional systems development by adopting a shift from rationalism to empiricism based on the data-learning paradigm. AI systems can hardly be understood and analyzed as the composition of elements. Neural networks are black boxes whose behavior and properties cannot be deduced from knowledge of the behavior of their elements. Although we know the functions computed by their elements, it is virtually impossible to characterize their input-output behavior by a mathematical model that we can analyze and reason about. As we have explained, the development of an AI system does not follow traditional step-by-step flows aimed at satisfying the properties implied by the system's requirements. It adopts a holistic approach aimed at discovering meaning and approaching the relationships encoded in complex multidimensional data. It consists of activities that start with data acquisition and preparation and lead to deployment, followed by system evaluation and improvement.

For all these reasons, AI systems cannot be verified and only in some cases can be tested when their input output behavior can be rigorously characterized. This is not possible when input or output data are sensory or linguistic data. AI systems also suffer other limitations such as anomalies e.g. non-robustness and adversarial attacks. As a result, AI systems are prone to unpredictable failures at runtime. To reconcile reliability requirements with the services of intelligent systems, one solution is to control their behavior using a Runtime Assurance (RTA) Architecture, as shown in Figure 14 [22, 3].

The Figure 14 shows an AI System that controls a network Facility so as to achieve a desired overall behavior. The Facility can be a fairly complex system, for which we want to achieve performance and adaptability that only an AI system can provide. The RTA is used to disconnect the control flow in the event of hazard of the AI System. It integrates a Trusted Monitor, a Trusted System and a Switch that so that the hazards detected by the Monitor trigger the Switch, which transmits control to the Trusted System. More precisely, the role of the components of the RTA is as follows.

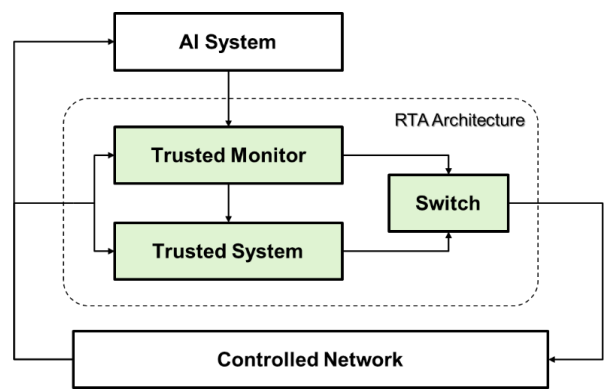

Figure 14 The RTA architecture illustration

*The Trusted Monitor* detects hazards as deviations from the desired behavior described by a set of critical properties. It must therefore be able to analyze and predict the impact of AI-generated results on the Facility in order to detect potential risks.

*The Trusted System* can cope with hazards detected by the Trusted Monitor with possible performance degradation, focusing on critical properties. As its role is to produce reliable decisions, it must be designed and implemented in such a way as to fully meet this objective. While it may sacrifice some degree of adaptability and performance compared to the AI system, it serves as the final safeguard for network control and must therefore meet the highest standards of trustworthiness.

The **Switch** provides the output of the **AI System** if no violation of critical properties is detected.

RTA must ensure continuity of service. If the AI system fails, the trusted system takes over. In turn, when the AI system recovers, the Monitor re-establishes control of the installation. Continuity of service can raise significant technical issues concerning the speed and smoothness of the transition between the two systems.

## 3.2 AI for autonomous networks

With the rapid progress of AI, the question of how to effectively apply AI techniques to key use cases in various sectors to improve efficiency and performance has become increasingly important. As previously noted, many use cases within telecommunications networks are highly critical and cannot tolerate errors. Data-driven AI, by its nature, introduces uncertainty and lack of explainability, which can lead to failures that are difficult to analyse and manage. Consequently, the application of AI in telecommunications needs to be approached with caution. However, there is no doubt that AI is essential to enable autonomy. Therefore, identifying and developing AI techniques suitable for autonomous networks represents a major challenge that requires further research.

### 3.2.1 Application of AI techniques to AN

From the outset, we need to rigorously classify the different types of AI to be applied in telecommunications, based on their intended purposes and corresponding use cases, as shown in Figure 15.

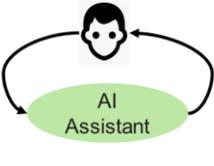

| | 1. Conversation | 2. Analysis | 2. Prediction | 3. Autonomy |
|---|---|---|---|---|
| Purpose | Human machine interaction | Continuously monitor a system and analyze its behavior | Continuously monitor a system and predict situations of interest in its behavior | Interact with environment showing awareness and acting in order to achieve goals |
| Technology | Generative AI (LLM) | Traditional ML or rule-based systems (symbolic) | Traditional ML or rule-based systems | Mixture of different types of AI |
| Use Cases | Human intent input, SLA report generation | RCA (root cause analyses), causal clustering and detection of relevant situations (classification, recommendation) | Prediction of potential safety or security issues and of performance variations | Decision making and action control to replace humans operators |

Figure 15 Different uses of AI in autonomous networks

➢ **For conversation**. AI assistants are designed to operate in question-and-answer mode in interaction with a human operator. These systems are set to play an important role in Human-Agent Interaction models by providing on-demand services such as SLA report generation, document summarization, configuration suggestions and so on. However, to facilitate effective communication with operation and maintenance (O&M) engineers, a major technical challenge is to enable AI assistants to understand network context and its relationship with linguistic data. To do this, they need to be able to learn and integrate the terminology, concepts and dynamics specific to telecoms networks into their language processing capabilities. Although AI assistants improve efficiency and reduce the level of expertise required for O&M tasks, humans continue to play the primary role in these operations. It is important to note that, despite the growing popularity of large language models and generative AI technologies, their integration into industrial applications currently remains limited.

➢ ***For analysis and prediction***. AI monitors are designed to continuously monitor system data and behavior to analyze or predict critical situations. These monitors are capable of processing complex data to uncover hidden insights. A typical use case in network management involves identifying and locating the root causes of alarms. AI monitors should be able to analyze alarm data over a specified period, considering factors such as occurrence times, locations, related logs, and KPI data, to accurately pinpoint the root cause. In the realm of prediction, a common use case is traffic forecasting, where AI monitors predict network traffic based on historical data. This enables proactive adjustments to network parameters, optimizing the user experience.

Building AI monitors presents significant challenges, particularly in selecting the appropriate AI techniques that align with the unique characteristics of the use case and data. LLMs are not suitable for building monitors, which are streamers that receive data streams and produce knowledge. In fact, we know that the probability of an erroneous response increases exponentially with the duration of the interaction. Therefore, for monitors, we should consider solutions based on traditional machine learning or rule-based systems.

AI monitors do not directly control networks; instead, they provide analysis and prediction capabilities that can replace certain human tasks. To a certain extent, they can be used to improve the autonomy of O&M operations.

➢ ***For autonomy***. AI agents are attracting increasing interest and are now becoming a central topic of artificial intelligence research. However, in the current state of knowledge, they cannot be used to fully implement network agents. They can be entrusted with non-critical tasks such as collecting and synthesizing information on the state of the network and its services, and finding solutions to problems such as navigating through more or less static structures. The use of information generated by AI agents without filtering by human experts should only be considered if their decisions do not affect critical network functions. For example, to optimize parameters whose values do not have a critical impact.

The application of AI techniques to AN is facing several significant challenges as below.

➢ ***Techniques for telecommunication network data***. Relevant techniques include data classification, data cleaning, the development of training datasets, and the validation and evaluation of the obtained solutions. Obviously, data should reflect not only relations between measured quantities but also topological relations induced by the structure of the networks. The training dataset should encompass various types of data, including topology, KPIs, packet attributes, logs, and the effect of actions on network states. Its development will require human expertise and analysis techniques to assess the relevance of the various data fields in relation to the types of knowledge to be produced. For instance, the development of an AI alarm analyzer to identify root hazard causes requires that the AI system takes into account the types of the alarms and their position in the network hierarchy. Otherwise. It will not be able to provide diagnostics and localize root causes.

➢ **Training LLMs to build a Copilot**. LLMs have been successfully applied to process multi-modal media data, such as images, videos, text, and speech. Network data is also multi-modal, but the nature of network data differs significantly from that of media data, which may require specific embedding techniques. The challenge is to integrate this heterogeneous data into a common semantic domain, while ensuring that essential information is preserved throughout the process. To learn effectively from multimodal network data, an appropriate LLM model must be developed, capable of aligning domain-specific information in its internal parameter space and accurately interpreting the network context. It must be able to consider technical information about the network, its structure and state, and to predict to some extent the effect of applicable actions.

➢ **Transferring O&M expertise into knowledge**. It is a central issue for the development of AI components used in the architecture. This task will be greatly facilitated by the development of a specific data infrastructure based on ontologies and a rigorous definition of world knowledge, as explained in the next section. However, the difficulties encountered in successfully transferring human expertise to expert systems give some idea of the difficulty of the undertaking.

Below, we look at trends in the creation of AI-enabled agents in so-called agentic AI or world models. These agents only partially cover the functionalities of our reference architecture.

### 3.2.2    Agentic AI and World Models

The term "agentic AI" [23] is used for systems that work in simple (essentially static or changing slowly) digital environments with human interaction sending commands such as "answer my e-mails", "find documents relevant to the battle of Waterloo" etc. These agents are not constrained dynamically and their goals are simply obtained by direct interpretation of the user queries. They could be used in AN to perform background tasks that have no direct impact on the behavior of the network, but which collect and provide information to a human operator. Co-pilots could integrate this type of agents, in addition to conversational functionality.

The term "World Model" [24] refers to agents whose behavior is primarily reactive, as they interpret an environment that can change dynamically. The input stimulus is usually visual information from a physical or virtual environment. A World Model estimates the state of the environment and predict future actions, partially covering the reactive behavior of the architecture. However, its ability to make in-depth predictions and achieve objectives is limited. Some work suggests the possibility of making predictions by simulation, but this does not seem realistic, as it presupposes the availability of a model of the agent's interactions with its environment.

Although there is a consensus that World Models should integrate symbolic reasoning capabilities, there is no agreement about how this can be achieved.

One approach considers that reasoning capability can be achieved using only data-driven techniques. In other words, reasoning will emerge as we build increasingly complex machine learning systems. This approach naturally leads to the development of end-to-end autonomous AI agents, as some players in the automotive industry are already attempting to do [25, 26].

The other approach considers that some reasoning and explicit knowledge must be part of the world model from the outset. This leads us to consider agent architectures where the AI components use symbolic engines [27] or are supplemented by knowledge of long-term memory, as suggested by the RAG architecture paradigm [28], as shown in Figure 16. The latter are considered as a promising approach for improving precision and mitigating hallucinations of LLMs.

Figure 16 shows an LLM receiving a prompt that is a percept obtained by analysis of sensory information from the agent's environment. The LLM provides a first response to this prompt and sends it to a retrieval mechanism. For instance, if the prompt is "I would like to know about the state of the links of the network" the LLM will generate a set of queries aiming at acquiring detailed information about state parameters and their inter-relations. The retrieval mechanisms will explore the Memory and extract all the relevant information (Best Results). This information may be detailed is sent back to the LLM that will further analyse and summarize it to provide the best estimate of the state and predict possible actions.

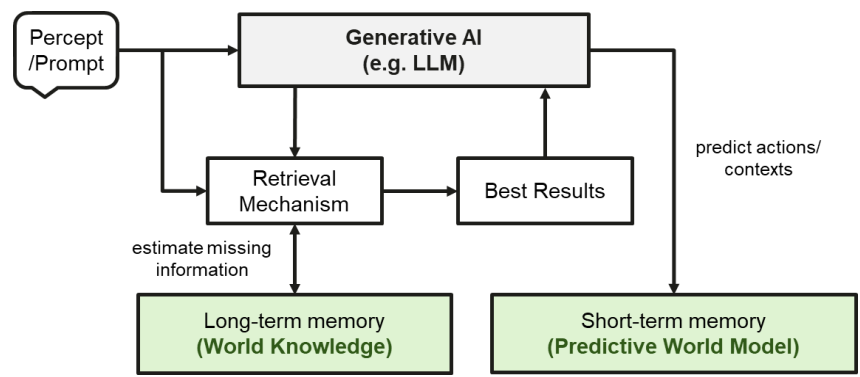

Figure 16 RAG Architecture integrating Generative AI & World Models

The construction of efficient RAG architectures would be a major step towards the realization of autonomous agents. Indeed, the realization of our agent Reference Architecture requires the application of this paradigm to implement the interactions of the modules with the Memory that contains World Knowledge.

Note that RAG architectures raise problems not directly related to machine learning and have to do with extraction of information relevant to a query according to some similarity criteria that should implicitly take into account semantics. Without getting into details, there are different ways to encode and store knowledge in the Memory, e.g. by indexing, using vectors or semantic graphs. Obviously, the retrieval mechanism and the associated similarity relation should take into account the encoding used.

### 3.2.3 Agent properties

Rigorous specification of properties is essential for expressing requirements and system validation. The behavior of AI systems, insofar as they seek to mimic that of humans, can be characterized by human-centric properties that are very different from the technical properties of traditional systems. Many works emphasize the importance of these properties focusing on "responsible", and "ethical" AI or AI aligned with human values. However, most of this work lacks a foundation, as the definition of these properties requires their connection to a semantic model on which they could be evaluated to decide on their validity. We explain that our agent reference architecture could be a suitable model provided we have access to World Knowledge.

The properties are declarative sentences expressing the relationships between their world states and their evolution. They are built from a set of atomic state predicates using logical operators. Given the meaning of the atomic predicates, it is possible to attribute a meaning to the property. The agent state defines a valuation of the atomic predicates of a property, and consequently its logical value, true or false.

An important difference between traditional systems and AI systems is that, in addition to the usual behavioral properties, they can be asked to satisfy cognitive properties.

➢ **Behavioral properties**, can be specified using predicates based on the observed behavior of the input/output system, independently of the system's other structural and physical properties. All the essential properties of traditional systems are behavioral properties. These properties can also be defined for AI systems with the difference that their validity may not be asserted based on objective criteria.

➢ **Cognitive properties**, depend mainly on the system knowledge and self-awareness characterized by its reasoning capabilities, as well as its decision and problem-solving process. As explained below, these properties are important for understanding the extent to which AI systems are aligned with human values.

| System operation / Requirements | Behavioral Properties | Cognitive Properties |
|---|---|---|
| Risk-related properties | Safety & Security | Normative (ethical/law-enforced) |
| Usefulness properties | Functionality, Performance, Efficiency, and User-friendliness | Purpose-oriented, Goal-oriented, and Rationality properties |

Figure 17 The distinction between behavioral and cognitive properties

Figure 17 shows a classification of a system's properties according to its requirements and the two main aspects of its behavior. Requirements distinguish between risk-related and usefulness properties. The former expresses the system's resilience to any type of hazard caused by various risk factors. The latter express the extent to which the system provides the expected functionality and service. This distinction divides cognitive properties into two categories.

On the one hand, ethical and legal properties, which require that a system's knowledgeable action conforms to rules that prevent it from endangering itself or its environment. On the other hand, the properties that characterize systems enable us to identify the actions motivated by its purpose and objectives and driven by the system's rational thinking.

Note that to decide cognitive properties, we need to access the agent's World Knowledge, as well as the way in which intent is manifested and processed by the agent's proactive part. In fact, we trust humans because we know that they know the rules of a value system and are bound to respect them. For example, it is impossible to decide whether the agent is acting ethically if we do not know whether it is aware of the ethical rules in force. The specification and validation of cognitive properties are still poorly understood and deserve further investigation.

## 3.3   World Knowledge

### 3.3.1   The concept of world knowledge

| Knowledge | | | |
|---|---|---|---|
| Cognitive and Deontic | **Agent-specifc** (subjective) | **General** (objective) | Scientific and Technical |
| Implicit empirical, e.g. mathematical and scientific theories Properties, declarative rules | **Declarative** | **Procedural** | Empirical methods, e.g. design and system development flows, coordination principles, imperative rules |
| Implicit human knowledge, e.g. machine learning knowledge | **Data-based** | **Model-based** | Explainable by models e.g. formal or simply conceptual |
| Objects, physical or artefacts agents, natural or artificial | **Concrete concepts** | **Abstract concepts** | A posteriori e.g. scientific and technical A priori e.g. mathematics, logic Self-concepts e.g. intention, belief, feeling |

Figure 18 Types of world knowledge

In the architecture, World Knowledge refers to the information that an agent has about the world around it and itself, which is used for decision-making and choice of goals. A systematic classification of World Knowledge, shown in Figure 18, according to its degree of validity, its generality and mode of use, is very important for its proper application by the agent.

➤ *Agent-specific knowledge and global knowledge*.

- *Agent-specific knowledge* is particular to an agent, generated from local data and reflecting what the agent individually knows or believes. For example, an agent may consider a network link to be active on the basis of its local knowledge, whereas the link is broken.

- *General knowledge* shared by all agents in the network, which must be coherent and valid. Indeed, one of the objectives of the interaction between agents is to reach a consensus on the creation and sharing of global knowledge.

➤ *Declarative and procedural knowledge*

- *Declarative knowledge* refers to statements of facts about the world that specify a large number of possible behaviors. It is often stored in the form of relations in databases or ontologies, ranging from simple facts to symbolic knowledge in the form of properties, or axiomatic systems formalizing the knowledge of a domain.

- *Procedural knowledge* refers to knowledge relating to methods for calculating functions, executing tasks or achieving objectives in the form of a sequence of steps.

➤ *Data-based and model-based knowledge*

- *Data-based knowledge* refers to implicit knowledge extracted from data using data analytics or machine learning techniques.

- *Model-based knowledge* is described by models whose semantics is rigorously defined and

which relate them to states of the agent's world. It is explainable and verifiable; for example, we can use decision methods to determine its veracity for a given state of the world.

➢ **Concrete and abstract concepts**

- *Concrete concepts* refer to real objects with observable and measurable attributes, such as an optical cable of a certain length.

- *Abstract concepts* are intangible and do not admit of concrete physical realization. They express mental qualities or relationships, in particular elements of models, used to reason about the world through abstraction and metaphor.

The formalization of different types of knowledge can be based on ontologies, which support a rigorous methodology for specifying concepts and their relationships in a hierarchical way, for given domains. Ontologies have been used in systems engineering in a wide range of applications, for the specification of system requirements, properties and behavior.

### 3.3.2    World Knowledge for AN

In the context of an autonomous network, world knowledge refers to how the AN agent perceives the network environment and the agent state. The knowledge is used to support predicting network state, making network decisions, choosing suitable goals, providing O&M suggestions to humans, and collaborating with other agents. It can be further refined as general and specific network knowledge, as explained in Section 2.2.6.

To make an agent deployed in the network as autonomous as possible, the biggest challenge is to effectively transfer human O&M expertise to the agent's knowledge, either through machine learning or symbolic techniques. To achieve this, we need to 1) define and describe the agent's O&M knowledge, and 2) progressively develop and assemble the building blocks that are sorely lacking to realize the architectural components. We see three technical challenges that require further exploration.

➢ **Transforming O&M expertise into rules-based knowledge.** O&M expertise and empirical knowledge can often be formulated using rules expressed in the form of condition-action pairs, based on a limited set of parameters. For example, if a network indicator exceeds a certain threshold (condition), a specific parameter should be configured (action). This type of knowledge is well-suited for formalization using symbolic AI techniques, particularly expert systems, which can translate these rules into logic for automated reasoning.

➢ **Data-based analysis and prediction**. Data-based techniques and machine learning in particular, can be used to generate knowledge for prediction and analysis, hidden in raw, unstructured data. As explained in the previous section, root-cause analysis is a typical example when a large number of alarms are triggered, O&M staff need to analyze various data sources, such as device logs, performance indicators and topology, in order to identify the root cause. This process is primarily empirical, as it concerns not only individual data points, but also the way in which these data points interact with each other in the wider context of the network. The challenge is to unify and align this information, as well as to understand the underlying relationships in the context of the network structure. Although AI technologies have been particularly effective in generating knowledge through learning from general data, their application in technical domains for predictive and analytical purposes remains a challenge. To be successful in analyzing these kinds of problems, the AI system used must have knowledge of the network and its graph-like structure, as well as of the possible relationships and interactions between its various components.

➢ ***Developing a domain-specific Copilot***. A co-pilot is an LLM that has access to World Knowledge, possibly using a RAG architecture, enabling it to retrieve relevant information and provide enhanced context for user queries. The main challenge is to develop an LLM capable of understanding network-related information. The LLM-driven co-pilot needs to be trained on a specific network dataset, enabling it to process both technical concepts and network actions. In addition, network-specific prompting techniques need to be developed to ensure that the co-pilot's responses are context-appropriate and useful to the engineers. Integrating these technical requirements to develop a telecom-specific co-pilot remains an open problem, with significant challenges in network modeling and information retrieval.

## 3.4   Achieving collective intelligence

An autonomous system is made up of agents, each of which pursues specific goals, but which must also coordinate to achieve the system's overall goals. Therefore, the correctness of an agent's behavior also depends on its ability to coordinate with other agents in such a way that its actions not only do not prevent other agents from fulfilling their goal, but also create the synergies needed to satisfy global properties. It is therefore essential to distinguish between the individual intelligence of an agent and the collective intelligence displayed by the agents of an autonomous system.

Figure 19 illustrates the difficulty of building autonomous systems, taking into account their situational awareness and decision-making capabilities. Situational awareness becomes more difficult as we move from a single domain to multi-domain knowledge and open world awareness. The difficulty of decision-making increases with the number of agents and the number of objectives to be managed.

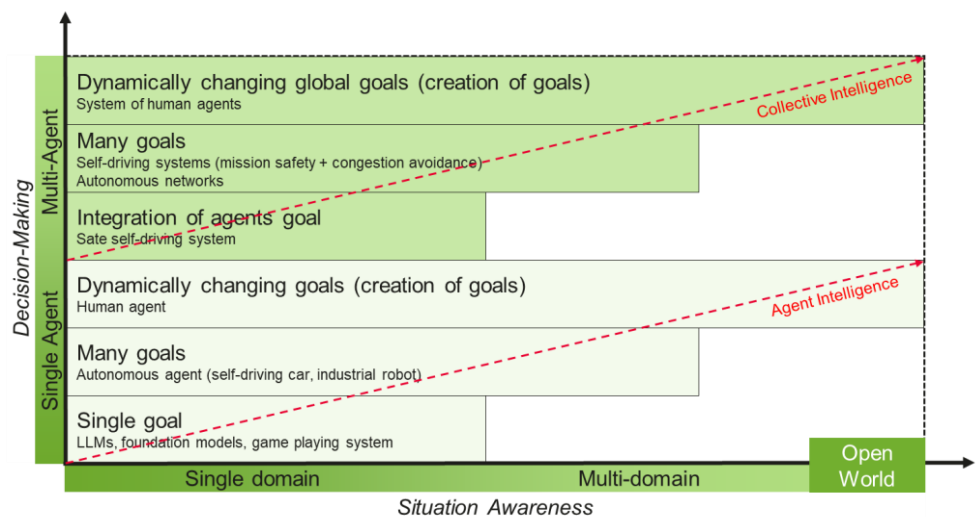

Figure 19 Agent Intelligence and Collective Intelligence

According to these criteria, conversational systems, such as LLMs, can process multimodal sensory information for a single purpose. However, their situational awareness is limited compared to human situational awareness, which relies on common sense. For autonomous vehicles, the demands on awareness are higher than what LLMs can achieve today.

As far as goal management is concerned, the difficulty increases as we move from conversational systems to autonomous systems managing numerous goals. In the latter case, an additional difficulty arises from the possibility of managing the creation of new goals.

We distinguish three levels of collective intelligence characterized by the set of global properties satisfied by the system.

➢ Safe integration of agents, i.e. interaction between agents does not prevent the achievement of each agent's objectives. Problems can arise when the system's coordination mechanisms are not free of blockages, or when the underlying mechanisms for resolving conflicts between agents are not fair.

➢ The agent system successfully manages a predefined set of global objectives through agent coordination, such as self-reconfiguration or self-repair. This requires appropriate distributed coordination mechanisms, such as protocols implemented by message exchange between the Agent-Agent Interaction modules of the system's agents.

➢ The system can dynamically create new global goals as needed, to adapt to changing situations in its environment. Only human organizations can demonstrate this level of collective intelligence.

Building an autonomous system with multiple agents raises difficult systems engineering problems that have nothing to do with agent intelligence.

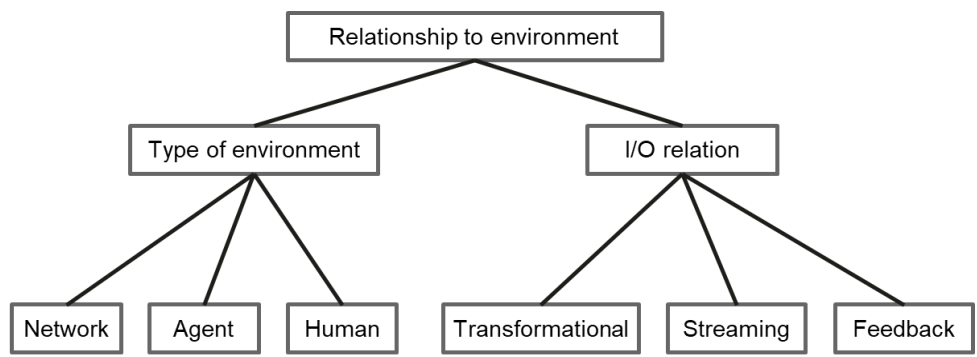

Figure 20 Agent integration in its environment

Some problems are linked to the integration of agents in their environment (Figure 15). There are three main types of environment for an agent: (1) the environment constituted by the network elements that the agent controls (2) the human environment constituted by the operation and maintenance (O&M) engineers and users with whom the agent must collaborate harmoniously. This collaboration requires appropriate interfaces to enable the O&M engineer to understand the situations to be managed and to control them in the event of intervention (3) the environment comprising all the agents with which coordination is required.

Other problems relate to the implementation of appropriate coordination mechanisms, as shown in figure 16. Agent coordination requires specific architectures, depending on the overall objectives to be achieved. The simplest type of architecture is one in which the number of agents and their interactions are fixed. Parametric architectures are coordination schemes that can be applied to any number of agents. For AN, we need dynamic architectures. We distinguish three types of dynamism. (1) Temporal dynamism, which means that the number of agents and their interactions can change over time, for example when agents and their interactions are created or deleted. (2) Spatial dynamism: agents' behavior changes according to their position in space. This is the case with mobile agents. (3) Organizational dynamism: agents change their behavior according to their position in the system's organization. The latter generally evolves dynamically.

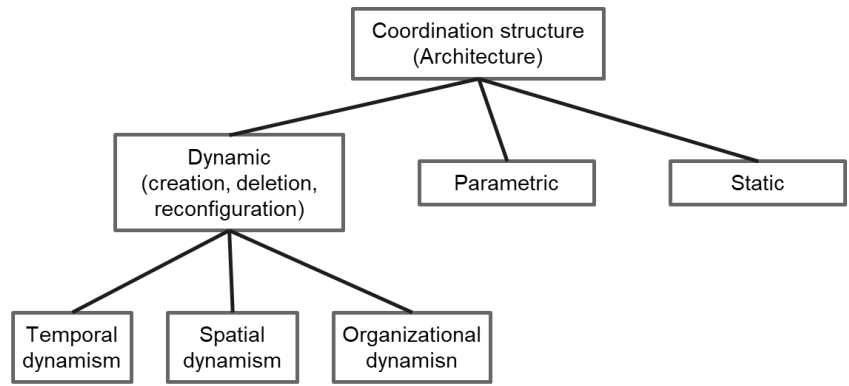

Figure 21 Complexity of coordination structure

For autonomous networks with hierarchically distributed agents, both temporal and organizational dynamism are required. Spatial dynamism is only necessary in certain network types and use cases, as most network devices are geographically fixed.

The above analysis shows that individual agent intelligence is not enough to build a correct autonomous network. It is easy to build problematic systems with highly intelligent agents that do not coordinate properly.

# 4   Concluding remarks

This document presents an overview of the key challenges designing autonomous networks and outlines a progressive architecture to bridge the gap from today's traditional approaches. By outlining a vision and the technical and economic requirements, including increased automation to reliably and efficiently replace O&M engineers, it develops the key idea of establishing a Reference Architecture, as a common basic model for their components. Such an idea is the basis of complex software system development practices. However, its application to autonomous systems raises several new questions, in particular, because these systems require the use of AI techniques to implement functions that are practically impossible to achieve with traditional software.

Autonomous networks are among the most difficult autonomous systems to build, insofar as they are distributed systems comprising a very large number of heterogeneous agents, each pursuing specific goals but which must cooperate harmoniously to demonstrate collective intelligence in order to satisfy global properties. A comparison of AN with other autonomous systems, such as self-driving vehicle, gives a better understanding of the scale of the challenge. The development of a self-driving system centers on a single agent interacting with mobile objects under well-defined traffic rules. In contrast, AN involve a variety of agents and their development must be holistic, balancing individual agent behavior with hierarchical and distributed interactions.

The properties of AN involve a wide variety of technical and economic constraints. The latter include time and performance constraints, as well as purely economic constraints regarding their operation and pricing of services. In order to satisfy these constraints, AN must satisfy global goals and thus demonstrate global intelligence.

The proposed reference architecture characterizes the composition of an autonomous agent as a set of basic functions specified by their input and output domains. We argue that the proposed architecture satisfies the properties of incremental scalability and completeness. We also show how the behavior of agents can be specialized in network use cases. Moreover, we present the instantiation guide of an agent according to its role and position in its hierarchical structure.

The reference architecture considers that an agent is integrated into the hierarchical environment of the AN by participating in three different types of interaction. 1) Interactions with the hierarchical structure downstream, of which it is the root and for which it acts as a controller; 2) Interactions with human operators who come from the upper part of the AN; 3) Interactions with other agents who are generally at the same layer.

Consequently, the architecture integrates two Human-Agent (HAI) and Agent-Agent (AAI) Interaction modules, as well as a set of modules generating the reactive behavior of the agent acting as a controller. Given the nature of the problem, the latter must be synchronous with the controlled environment and subject to time constraints in terms of delay and performance. In addition, the agent is equipped with a set of modules that generate proactive behavior aimed at dealing with problems that are managed in traditional networks, by human operators, such as counteracting any potential deviation of the observed network characteristics from requirements. In this way, proactive behavior complements HAI interaction by automating tasks that give rise to the intent to eliminate deviations with the emergence of new goals to be addressed by reactive behavior. The distinction between reactive and proactive behavior is a key feature of the proposed architecture that in a way reflects the distinction between System 1 and System 2 of human thinking [29].

Finally, a central element of the reference architecture is a long-term memory where knowledge used and updated by all the elements of the architecture is stored to improve and ensure the consistency of their predictive and decision-making capacities.

It should be emphasized that, in addition to the specification of the modules that make up the reference architecture, we also propose scenarios, which show how they coordinate to meet the requirements of autonomous behavior. While the architecture assigns distinct roles to modules, it allows for flexible implementations. Some agents may not need all the proposed functions. In addition, there may be implementations that combine a set of functions and their coordination. This flexibility is crucial in determining how much AI is needed to implementing the architecture's functions.

One can imagine, for example, end-to-end AI solutions that fully implement an agent. This type of solution does not seem viable in the current state of the art for several reasons, the most important being that current AI systems have relatively good capabilities for analyzing situations, but they have limited reasoning and action planning capabilities subject to real-time constraints in relation to dynamically changing environments. Furthermore, implementations without long-term memory where explicit knowledge is stored can be considered, using AI components that have been sufficiently well-trained with the necessary domain-specific knowledge.

As things currently stand, we can only consider hybrid solutions, which integrate traditional components with AI components, and for this, long-term memory seems to play an important role. In our analysis, we identify AI components that could provide the necessary functionalities in the architecture. For reactive behavior, World Models can cover some of its aspects but still seem to have weaknesses, as they are not reliable and deterministic enough to guarantee real-time constraints. In our opinion, current software solutions remain the only realistic choice.

We consider that in the HAI interaction module, we need two types of AI components. Co-pilots to support the decision-making process of human operators as well as AI components for the prediction and analysis of critical situations. We also believe that for greater efficiency, it is necessary to be able to combine data-based knowledge and symbolic knowledge. A certain type of human expertise can be directly specified by rules.

One of the great unknowns is the possibility of effectively using explicit long-term memory knowledge. While the idea of increasing the accuracy and robustness of analysis and decision-making systems by using explicit knowledge is consistent with the RAG paradigm, the current situation raises questions about the reliability of retrieval mechanisms based on the type of representation and associated similarity relationships.

In conclusion, our analysis shows that the road to AN will certainly be a long one, and the urgent question is what needs to be advanced in AI to accelerate progress towards achieving fully autonomous networks. A considerable effort must be made to develop an AI specifically trained to recognize not only basic technical concepts, but also to reason in terms of actions and goals. AI agent development is part of this, but we must be aware that these agents are still unreliable, and as such are only intended to provide non-critical services. There is an urgent need to develop AI components for the prediction and analysis of critical situations in networks. Domain-specific data infrastructure is necessary, and still needs to be developed to a large extent. It is impossible to imagine how we could obtain AI capable of performing root cause analysis in an alarm management system, without being led to recognize the network structures with their components and the type of alarms integrated into them.

Let us hope that our contribution sheds light on questions that are still the subject of much debate within the network community and contributes to a collective awareness with a view to a better understanding of the challenges and a concerted mobilization to resolve them. By addressing these challenges through collaborative research and engineering, we can accelerate progress towards truly autonomous networks.

# 5 Acknowledgement

The authors would like to thank Kevin McDonnell, Fernando Camacho, Lorenzo Cipriani, and Aleksandar Milenovic from Huawei's Irish Research Center, for their contribution to improving the document through constructive criticism and suggestions.

# Appendix A. Terminology for the AN agent reference architecture

This appendix provides a list of the modules and functions used in the AN agent reference architecture, along with their input and output data types.

| Term | Input | Output | Consist of |
|---|---|---|---|
| AN Agent | stimuli (from network environment); state / goal (from agent environment); goal / instruction / query (from human environment) | action (to the network environment); state / goal (from agent environment); linguistic data (to human environment) | Reactive Behavior; Proactive Behavior; World Knowledge; Human-Agent Interaction; Agent-Agent Interaction |
| Reactive Behavior | stimuli | action plan | Situation Awareness; Decision-Making |
| Situation Awareness | stimuli | predictive model | Perception; Reflection |
| Perception | stimuli | percept | - |
| Reflection | percept | predictive model | - |
| Decision-Making | predictive model | action plan | Goal Management; Planner |
| Goal Management | new goal predictive model | pursed goal | - |
| Planner | pursued goal | action plan | - |
| Proactive Behavior | agent state | new goal | Self-Awareness; Choice-Making |
| Self-Awareness | agent state | meta-goal (set of goals) | Agent Purpose; Intent Management |
| Agent Purpose | agent state | need | - |
| Intent Management | need; feasibility constraints | meta-goal | - |
| Choice-Making | meta-goal | new goal | Meta-goal Management; Choice of Goals |
| Meta-goal Management | meta-goal | goals | - |
| Choice of Goals | goals; feasibility constraints | new goal | - |
| World Knowledge | query (from other modules) | knowledge | Knowledge Manager; Knowledge Repository |
| Knowledge Manager | query | processed query | - |
| Knowledge Repository | processed query | retrieval knowledge | - |

| Human-Agent Interaction | goal / intent / action / query | linguistic data | User Interface; Problem Solver |
|---|---|---|---|
| User Interface | goal / intent / action / query | goal / intent / action to Knowledge Repository; query to Problem Solver | - |
| Problem Solver | query; knowledge | linguistic data | - |
| Agent-Agent Interaction | state; goal | updated state; updated goal | Global Awareness; Goal Coordination |
| Global Awareness | state of itself and from other agents | updated state | - |
| Goal Coordination | goal of itself and from other agents | updated goal | - |